\documentclass[aps,prb,twocolumn]{revtex4-1}

\usepackage[dvips]{graphicx}
\usepackage{epsfig}

\usepackage[margin=2cm,head=0.5cm]{geometry}
\usepackage[latin1]{inputenc}
\usepackage{bm}
\usepackage{amsmath}
\usepackage{amssymb}
\usepackage{latexsym}
\usepackage{amsfonts}
\usepackage{epsfig}
\usepackage{color}
\usepackage[linktocpage, colorlinks=true ,linkcolor=blue, citecolor=blue]{hyperref}
\usepackage[all]{hypcap}

\usepackage{framed} % to make frames around definitions and exercises...
\newenvironment{fshaded}{%
\MakeFramed {\FrameRestore}}%
{\endMakeFramed}

\newtheorem{defbox}{Box}
\newtheorem{Exercise}{Exercise}
\newtheorem{Solution}{Solution}

\newcommand{\expval}[1]{\left< #1 \right>}

\newcommand{\ket}[1]{\left|#1\right>}
\newcommand{\bra}[1]{\left<#1\right|}

\newcommand{\nn}{\nonumber\\}

\newcommand{\f}[1]{\mbox{\boldmath$#1$}}

\newcommand{\bea}{\begin{eqnarray}}
\newcommand{\eea}{\end{eqnarray}}

\newcommand{\trace}[1]{{\rm Tr}\left\{ #1 \right\}}
\newcommand{\ptrace}[2]{{\rm Tr_{#1}}\left\{ #2 \right\}}

\newcommand{\abs}[1]{{\left| #1 \right|}}

\newcommand{\ii}{\mathrm{i}}  % alternativ: {$\mathbf{i}$}  % imgaginäre Einheit

\begin{document}
  
\title{An electronic Maxwell demon in the coherent strong-coupling regime}

\author{Gernot Schaller}
\email{gernot.schaller@tu-berlin.de}
\author{Javier Cerrillo}
\author{Georg Engelhardt}
\affiliation{Institut f\"ur Theoretische Physik, Technische Universit\"at Berlin, Hardenbergstr. 36, D-10623 Berlin, Germany}
\author{Philipp Strasberg}
\affiliation{Physics and Materials Science Research Unit, University of Luxembourg, L-1511 Luxembourg, Luxembourg}

\begin{abstract}
We consider an external feedback control loop implementing the action of a Maxwell demon.
Applying control actions that are conditioned on measurement outcomes, the demon may transport electrons against a bias voltage
and thereby effectively converts information into electric power.
While the underlying model -- a feedback-controlled quantum dot that is coupled to two electronic leads -- is well explored in the
limit of small tunnel couplings, we can address the strong-coupling regime with a 
fermionic reaction-coordinate mapping.
This exact mapping transforms the setup into a serial triple quantum dot coupled to two leads.
We find that a continuous projective measurement of the central dot occupation would lead to a complete suppression
of electronic transport due to the quantum Zeno effect.
In contrast, by using a microscopic detector model we can implement a weak measurement, which allows for 
closure of the control loop without transport blockade.
Then, in the weak-coupling regime, the energy flows associated with the feedback loop are negligible, 
and dominantly the information gained in the measurement induces a bound for the generated electric power.
In the strong coupling limit, the protocol may require more energy for operating the control loop 
than electric power produced, such that the whole device is no longer information-dominated
and can thus not be interpreted as a Maxwell demon.
\end{abstract}
\date{\today}
\maketitle

%%%%%%%%%%%%%%%%%%%%%%%%%%%%%%%%%%%%%%%%%%%%%%%%%%%%%%%%%%%%%%%%%%%%%%%%%%%%%%%%%%%%%%%%%%%%%%%%%%%
%%%%%%%%%%%%%%%%%%%%%%%%%%%%%%%%%%%%%%%%%%%%%%%%%%%%%%%%%%%%%%%%%%%%%%%%%%%%%%%%%%%%%%%%%%%%%%%%%%%
%%%%%%%%%%%%%%%%%%%%%%%%%%%%%%%%%%%%%%%%%%%%%%%%%%%%%%%%%%%%%%%%%%%%%%%%%%%%%%%%%%%%%%%%%%%%%%%%%%%
%%%%%%%%%%%%%%%%%%%%%%%%%%%%%%%%%%%%%%%%%%%%%%%%%%%%%%%%%%%%%%%%%%%%%%%%%%%%%%%%%%%%%%%%%%%%%%%%%%%

\section{Introduction}

In the famous thought experiment, Maxwell's demon is an intelligent being that measures the direction and speed of particles in a box with two compartments.
By suitably opening or closing a shutter between the compartments, the demon can sort the initially thermally distributed particles into
cold and hot fractions.
This thermal gradient can be used to extract work.
Effectively, the feedback loop implemented by the demon leads to a {\em local} reduction of entropy, 
ideally without any energetic cost, only using the information from the measurement.
The possibility of converting information into work has inspired generations of researchers to investigate the role of information in thermodynamics~\cite{leff2002}.

With nowadays rapid improvement of competing experimental approaches, it has become possible to implement different 
versions of a Maxwell demon in real-world scenarios.
These approaches include electronic~\cite{koski2015a,chida2017a}, qubit-qubit~\cite{camati2016a}, qubit-cavity~\cite{cottet2017a}, 
and photonic~\cite{vidrighin2016a} implementations.
And beyond Maxwell's demon, which is concerned with the control of average currents, for electronic transport setups feedback schemes proposing 
the control of even higher moments~\cite{brandes2010a} have been experimentally implemented~\cite{wagner2016a}.
With such advanced experimental abilities, huge interest exists in exploring quantum implications of a Maxwell demon.

Generally, it should be noted that in the theoretical discussion of quantum feedback control devices, two fundamentally 
different approaches exist.
In an {\em autonomous} (also termed coherent or all-inclusive) feedback loop, the original quantum system is supplemented by another 
auxiliary quantum system, which modifies the dynamics to reach a specified objective.
This all-inclusive approach has the advantage of simpler balance equations for the joint entropy of system and controller.
However, such systems are hard to design for arbitrary feedback loops (both theoretically and experimentally) and are not very flexible as 
the feedback loop and thus the desired function is hard-wired in the device.
Alternatively, one can implement an {\em external} feedback loop by performing measurements on the quantum system, 
classically processing the obtained information, and performing conditional control operations on the quantum system 
just as in the original thought experiment.
By changing the classical control protocol, i.e., choosing different control actions, the scheme can be modified to achieve
different objectives.
In contrast to classical systems, which ideally remain unaltered by the measurement alone, the dynamics of quantum systems is 
modified already by a measurement, which can have drastic consequences such as the quantum Zeno effect.
Thus, while this second approach appears closer to the original setup and may be more flexible, it 
has the disadvantages that its theoretical discussion and experimental implementation are also demanding regarding
the inclusion of the quantum measurement process and the fidelity of measurement and control steps, respectively.
We note that it has been possible to relate the entropy balances of autonomous~\cite{strasberg2013a,horowitz2014a} 
and external~\cite{schaller2011b,esposito2012a} Maxwell demons with each other.
Furthermore, in electronic transport setups, both autonomous and external versions of Maxwell's demon have been 
experimentally implemented~\cite{koski2015a,chida2017a}.

Being introduced within the framework of classical physics, models discussing Maxwell's demon theoretically 
typically employ the weak coupling limit between the controlled system and its reservoirs~\cite{quan2006a,schaller2011b,averin2011a,bergli2013a}.
Here, the energy contained in the interaction between them is negligible.
By contrast, in the strong-coupling limit, it is known that this interaction energy is no longer 
negligible~\cite{schaller2013a,esposito2015a,esposito2015b,strasberg2016a,strasberg2017b,newman2017a,perarnau_llobet2017a}, 
and even the partition into system and reservoir components becomes less obvious.
In our model, this interaction energy is directly related with the energetic cost associated to opening or closing the shutter.
Therefore, it is an intriguing question how Maxwell's demon performs when the interactions between controlled system and its
reservoirs become strong. 

In this paper, we will attempt to discuss this case for an electronic external feedback loop.
Particularly, we aim at generalizing the setup in Ref.~\cite{schaller2011b} to the strong-coupling regime.
The generalization of the corresponding autonomous setup in Ref.~\cite{strasberg2013a} will be discussed in a 
companion paper~\cite{strasberg2018a}, but see also Ref.~\cite{walldorf2017a}.
On the technical side, we will employ a fermionic generalization of a reaction-coordinate mapping, which is frequently employed in bosonic 
systems~\cite{martinazzo2011a,iles_smith2014a,iles_smith2016a,strasberg2016a} to treat the strong-coupling and non-Markovian limit.
We will see that projective measurements will imply Zeno-related modifications~\cite{elouard2017a} to the Maxwell demon 
dynamics, which requires a generalized discussion of the control loop including weak measurements~\cite{wiseman2010}.
Since these methods are partially new, we will explicitly present them in the following.
Below in Sec.~\ref{SEC:model}, we briefly review the underlying model, discuss the fermionic reaction coordinate mapping to a triple quantum dot, and show how to set up the propagator
for single feedback cycles in case of strong (projective) and weak measurements of the central dot's occupation.
Afterwards, in Sec.~\ref{SEC:thermodynamics}, we discuss the thermodynamics by defining the heat currents and the energy injected by the measurement as well as the energy injected by the control.
We discuss the performance of the device in Sec.~\ref{SEC:results} before concluding.

%%%%%%%%%%%%%%%%%%%%%%%%%%%%%%%%%%%%%%%%%%%%%%%%%%%%%%%%%%%%%%%%%%%%%%%%%%%%%%%%%%%%%%%%%%%%%%%%%%%
%%%%%%%%%%%%%%%%%%%%%%%%%%%%%%%%%%%%%%%%%%%%%%%%%%%%%%%%%%%%%%%%%%%%%%%%%%%%%%%%%%%%%%%%%%%%%%%%%%%
%%%%%%%%%%%%%%%%%%%%%%%%%%%%%%%%%%%%%%%%%%%%%%%%%%%%%%%%%%%%%%%%%%%%%%%%%%%%%%%%%%%%%%%%%%%%%%%%%%%
%%%%%%%%%%%%%%%%%%%%%%%%%%%%%%%%%%%%%%%%%%%%%%%%%%%%%%%%%%%%%%%%%%%%%%%%%%%%%%%%%%%%%%%%%%%%%%%%%%%

\section{Model}\label{SEC:model}

\begin{figure}[ht]
\includegraphics[width=0.45\textwidth,clip=true]{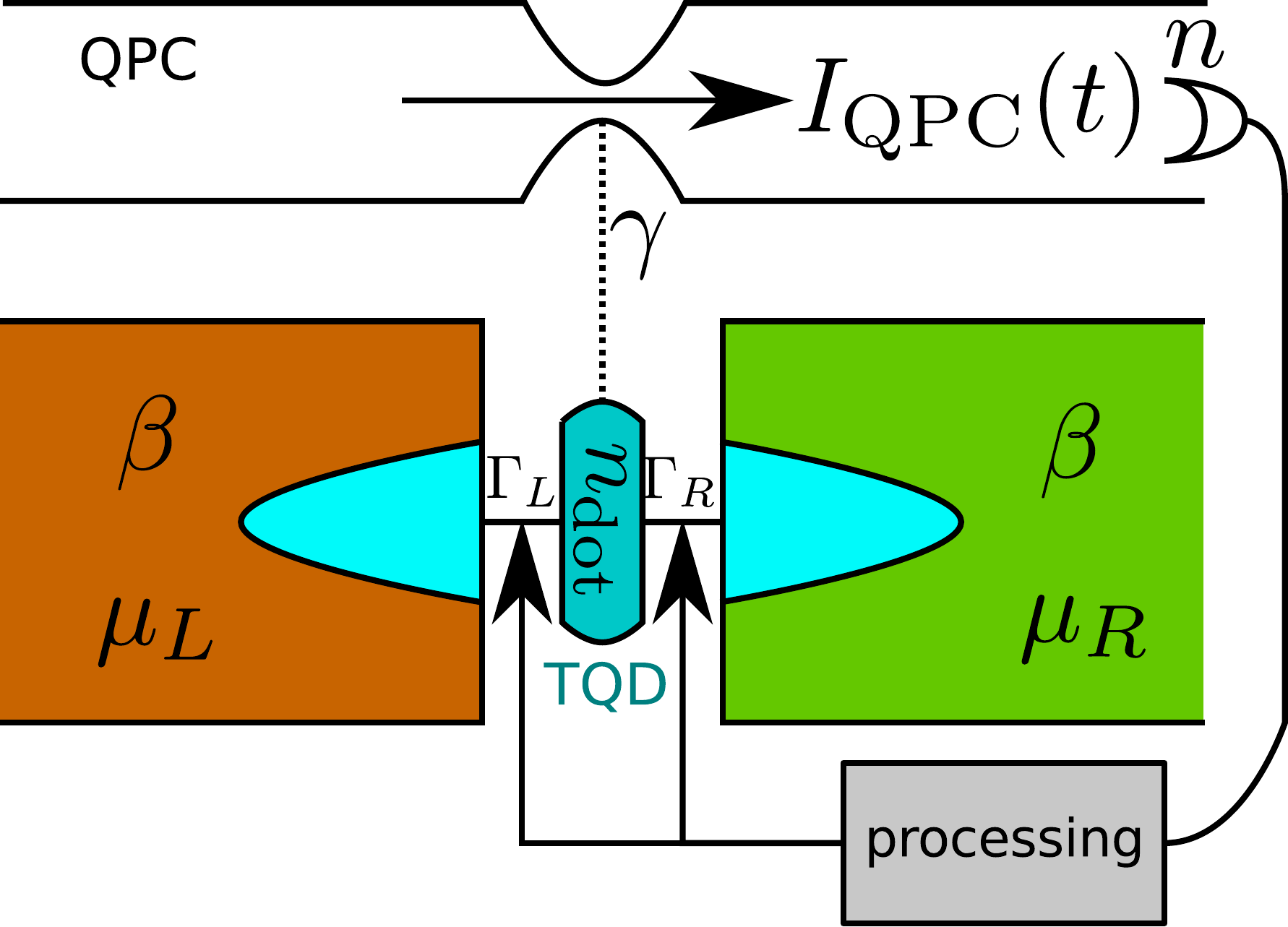}\\
\vspace{0.5cm}
\includegraphics[width=0.45\textwidth,clip=true]{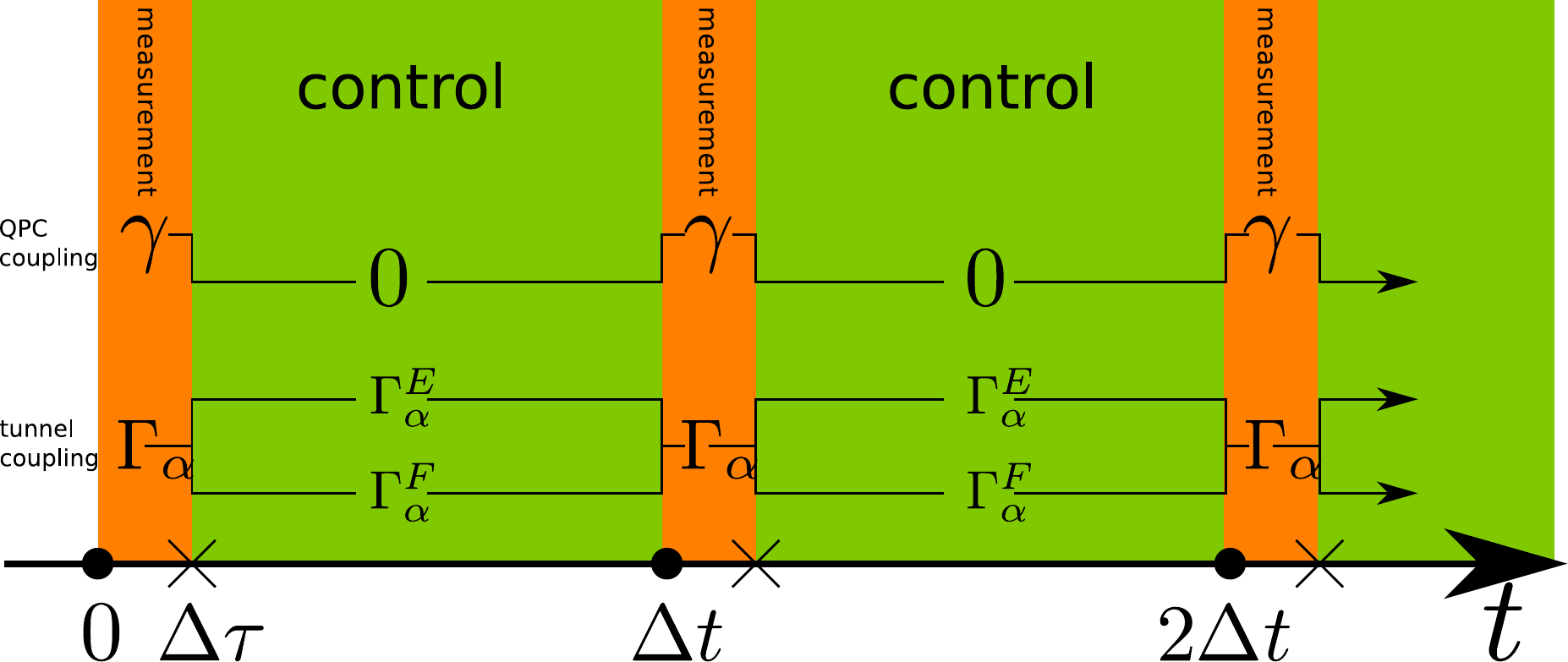}
\caption{\label{FIG:feedback_setup_sketch}
{\bf Top:} Sketch of the considered electronic transport setup with some relevant parameters.
Through the bottom circuit, the rates $\Gamma_\alpha$ allow for a current flowing 
between the left and right leads characterized by the inverse temperatures $\beta$ and the chemical potentials $\mu_\alpha$.
Via monitoring the occupation $n_{\rm dot}$ of the central dot with a nearby quantum point contact (QPC), the QPC current signal 
(obtained by measuring $n$ QPC charges in time interval $\Delta \tau$) can
be fed back into the system by conditioning the tunneling rates on the measured dot occupation.
We aim at the strong coupling limit by including collective reservoir degrees of freedom into the system (blue region).
{\bf Bottom:}
Schedule of the feedback cycles composed from repeated but alternating application of measurement phases (orange) of duration $\Delta\tau$, followed by
control phases (green) of duration $\Delta t-\Delta\tau$.
In this paper, we consider the limit $\Delta\tau\to 0$, keeping however the effect of the measurement ($\gamma\Delta\tau$) finite.
During control (implemented by conditional rates $\Gamma_\alpha^{E/F}$), the QPC detector is formally decoupled 
and the system Hamiltonian is conditioned on the preceding measurement, 
and during measurement, the system Hamiltonian is fixed to $H_{\rm S}$ (implemented by fixed rates $\Gamma_\alpha$), 
and the QPC interaction is dominating the central dot dynamics.
}
\end{figure}

In this section, we first briefly review the original model system  
in presence of projective measurements and piecewise-constant feedback control in Sec.~\ref{SEC:singledot}.
Then, we show how to map it to an equivalent triple quantum dot model in Sec.~\ref{SEC:tripledot}, 
where a Markovian embedding in an extended space allows to treat non-Markovian and strong-coupling effects in the original system.
Afterwards, we discuss the effect of a projective (strong) measurement on such a triple quantum dot in Sec.~\ref{SEC:projective}
and show that to avoid Zeno blocking, the introduction of weak measurements is necessary.
Finally, we discuss the weak-measurement implications of a microscopic detector model in Sec.~\ref{SEC:weak}, with
technical details exposed in App.~\ref{APP:measurement}, and close the feedback loop in Sec.~\ref{SEC:feedback}.
For orientation, we depict the setup and feedback cycle we have in mind in Fig.~\ref{FIG:feedback_setup_sketch}.

%%%%%%%%%%%%%%%%%%%%%%%%%%%%%%%%%%%%%%%%%%%%%%%%%%%%%%%%%%%%%%%%%%%%%%%%%%%%%%%%%%%%%%%%%%%%%%%%%%%
%%%%%%%%%%%%%%%%%%%%%%%%%%%%%%%%%%%%%%%%%%%%%%%%%%%%%%%%%%%%%%%%%%%%%%%%%%%%%%%%%%%%%%%%%%%%%%%%%%%

\subsection{SQD with projective measurements}\label{SEC:singledot}

The system we are aiming to control is a single electron transistor, where a single quantum dot (SQD) is tunnel-coupled to two leads
\bea
H = \epsilon d^\dagger d + \sum_{k\alpha} \left(t_{k\alpha} d c_{k\alpha}^\dagger + {\rm h.c.}\right)
+ \sum_{k\alpha} \epsilon_{k\alpha} c_{k\alpha}^\dagger c_{k\alpha}\,.
\eea
Here, $\epsilon$ denotes the dot level and $d$ the annihilation operator of an electron in the dot, which can tunnel via the amplitudes $t_{k\alpha}$ into 
left or right reservoirs $\alpha\in\{{\rm L},{\rm R}\}$, described by non-interacting modes with annihilation operators $c_{k\alpha}$ and energies $\epsilon_{k\alpha}$.
Spin effects are not considered throughout this work (spin-polarized electrons).
In absence of feedback, the single electron transistor is exactly solvable for arbitrary coupling strengths, see e.g. Ref.~\cite{topp2015a}.
However, in the weak-coupling limit a simple rate matrix ${\cal W} = {\cal W}_{\rm L} + {\cal W}_{\rm R}$
is sufficient to describe the evolution $\dot{\f{P}} = {\cal W} \f{P}$ of the probabilities $\f{P} = (p_{\rm  E}, p_{\rm  F})^{\rm T}$ for the empty and filled dot state 
\bea\label{EQ:rateeq_set}
{\cal W}_\alpha = \Gamma_\alpha^{(0)}(\epsilon) \left(\begin{array}{cc}
-f_\alpha(\epsilon) & +[1-f_\alpha(\epsilon)]\\
+f_\alpha(\epsilon) & -[1-f_\alpha(\epsilon)]
\end{array}\right)\,.
\eea
Here, $f_\alpha(\epsilon) = \left[e^{\beta_\alpha(\epsilon-\mu_\alpha)}+1\right]^{-1}$ denotes the Fermi function
of reservoir $\alpha\in\{L,R\}$ with chemical potential $\mu_\alpha$ and inverse temperature $\beta_\alpha$, and
the overall prefactor is determined by the spectral coupling density (SD)
\bea
\Gamma_\alpha^{(0)}(\omega) = 2\pi \sum_k \abs{t_{k\alpha}}^2 \delta(\omega-\epsilon_{k\alpha})\,,
\eea
which contains the non-thermal reservoir properties such as its level distribution $\epsilon_{k\alpha}$ and the coupling strengths $t_{k\alpha}$ of
individual reservoir modes $k$ to the system.
By keeping the reservoirs at different temperatures and chemical potentials, the total system can be interpreted as a 
thermoelectric generator~\cite{esposito2009b}, where a thermal gradient can be harnessed to drive electronic
transport against a voltage gradient.
Now, when additionally placing a quantum point contact (QPC) near the quantum dot, it is possible to read out the time-dependent dot occupation with high precision~\cite{flindt2009a}.
The current signal can be processed and fed back into the system by changing the tunneling rates in a time-dependent fashion.
In a simplified treatment, this can be treated as a sequence of instantaneous projective measurements, followed by a period of conditional piecewise-constant evolution, where
$\Gamma_\alpha^{(0)}(\epsilon) \to \Gamma_\alpha^{{\rm E}/{\rm F}}$ depending on the measurement outcome empty/filled, respectively.
After averaging over all outcomes and considering the limit of continuous measurements~\cite{schaller2012a}, one obtains an effective rate matrix under feedback
\bea\label{EQ:ratematrix_feedback}
{\cal W}_{\alpha} = \left(\begin{array}{cc}
-\Gamma_\alpha^{\rm E} f_\alpha(\epsilon) & +\Gamma_\alpha^{\rm F} [1-f_\alpha(\epsilon)]\\
+\Gamma_\alpha^{\rm E} f_\alpha(\epsilon) & -\Gamma_\alpha^{\rm F} [1-f_\alpha(\epsilon)]
\end{array}\right)\,,
\eea
which breaks the conventional local detailed balance relation.
Even in absence of a temperature gradient this can be used to generate electric power~\cite{schaller2011b}, which
can be interpreted as an electronic Maxwell demon.
We stress that a modified version of the fluctuation theorem is still valid~\cite{schaller2011b}, and 
the particular form of broken detailed balance implies that the second law is obeyed when the
information current resulting from the feedback loop is included in the entropic balance~\cite{esposito2012a}.
Very recently, this feedback scheme has been experimentally verified~\cite{chida2017a}.

However, there are some limitations in the SQD treatment.
First, the scheme is valid in the weak-coupling limit only $\beta_\alpha \Gamma_\alpha^{{\rm E}/{\rm F}} \ll 1$, such that for strong feedback driving the results should be questionable.
Second, the discussion of this scheme as a Maxwell demon lacks the calculation of the energetic balance done 
with the control actions, since this is neglected in the conventional master equation treatment.
Third, the treatment with a projective measurement does not fully comply with the experimental situation.
With the present contribution, we would like to overcome these limitations.

%%%%%%%%%%%%%%%%%%%%%%%%%%%%%%%%%%%%%%%%%%%%%%%%%%%%%%%%%%%%%%%%%%%%%%%%%%%%%%%%%%%%%%%%%%%%%%%%%%%
%%%%%%%%%%%%%%%%%%%%%%%%%%%%%%%%%%%%%%%%%%%%%%%%%%%%%%%%%%%%%%%%%%%%%%%%%%%%%%%%%%%%%%%%%%%%%%%%%%%

\subsection{TQD without measurements}\label{SEC:tripledot}

\begin{figure}[ht]
\includegraphics[width=0.45\textwidth,clip=true]{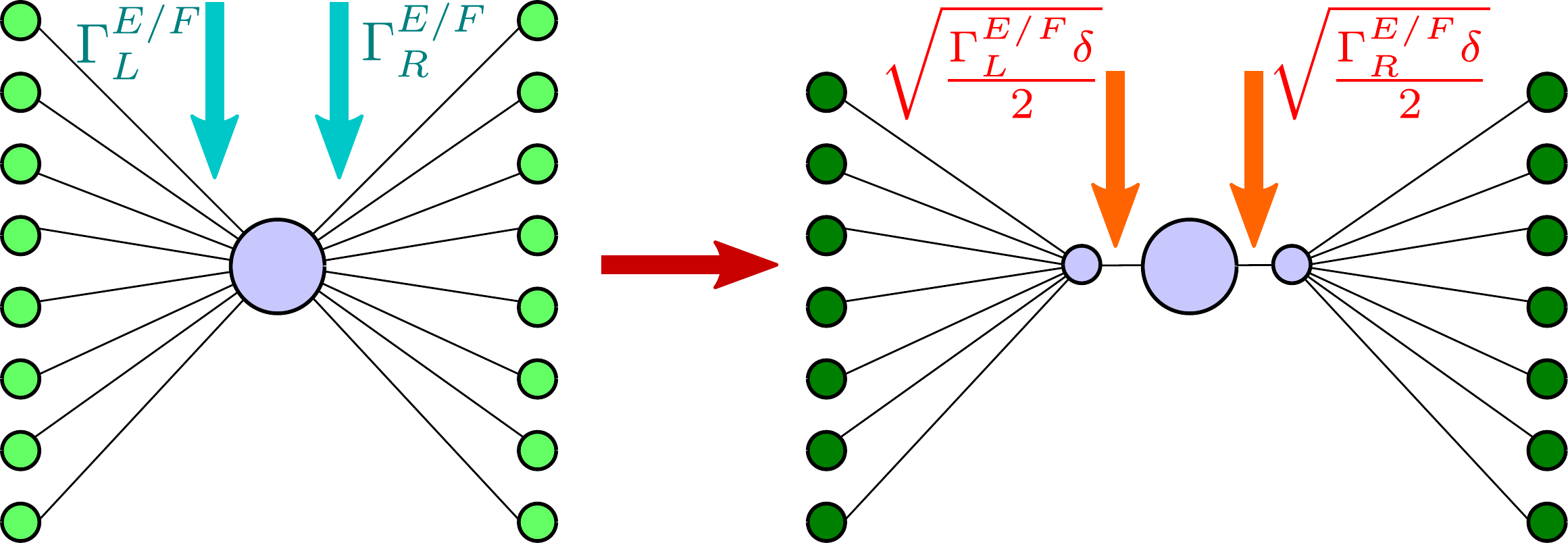}
\caption{\label{FIG:rcsketch_set_wfb}
Sketch of the reaction coordinate mapping from the SQD (left) to the TQD (right) model. 
A collective degree of freedom is separated from each reservoir of the SQD and absorbed into a redefined system, the TQD, which
is still tunnel-coupled to two residual reservoirs.
The original feedback loop on the SQD that modifies the tunneling rates in a Markovian treatment is thereby mapped to 
a feedback loop on the TQD, where the internal tunneling amplitudes within the TQD are changed in a piecewise-constant fashion.
}
\end{figure}

By applying separate fermionic Bogoliubov transforms for each reservoir, we can include separate reaction coordinates into the system, 
which maps our setup to a serial triple quantum dot system (TQD) that is tunnel-coupled to two residual reservoirs
via the renormalized tunneling amplitudes $T_{k\alpha}$, see Fig.~\ref{FIG:rcsketch_set_wfb} for an illustration.

After the mapping, the Hamiltonian of the TQD assumes the form
\bea\label{EQ:ham_mapped}
H &=& \Omega_{\rm L} d_L^\dagger d_{\rm L} + \epsilon d^\dagger d + \Omega_{\rm R} d_{\rm R}^\dagger d_{\rm R}\nn
&& + \lambda_{\rm L} \left(d_{\rm L}  d^\dagger + d d_{\rm L}^\dagger\right)
+ \lambda_{\rm R} \left(d_{\rm R} d^\dagger + d d_{\rm R}^\dagger\right)\nn
&& + \sum_\alpha \sum_k\left(T_{k\alpha} d_{\alpha} C_{k\alpha}^\dagger + T_{k\alpha}^* C_{k\alpha} d_{\alpha}^\dagger\right)\nn
&&+ \sum_\alpha \sum_k \tilde{\epsilon}_{k\alpha} C_{k\alpha}^\dagger C_{k\alpha}\,.
\eea
Here, the first two lines denotes the TQD system Hamiltonian $H_{\rm S}$ with reaction-coordinate on-site energies $\Omega_\alpha$ and TQD internal tunneling
amplitudes $\lambda_\alpha$, the third line contains the coupling, and the last line the residual reservoir terms.
Based on this TQD model, a new SD can be introduced
\bea
\Gamma_\alpha^{(1)}(\omega) = 2\pi \sum_k \abs{T_{k\alpha}}^2 \delta(\omega-\tilde{\epsilon}_{k\alpha})\,,
\eea
which can be obtained from the original SD with complex calculus methods.
More details regarding the fermionic reaction coordinate mapping are exposed in App.~\ref{APP:reaction_coordinate} and a companion paper~\cite{strasberg2018a}.
Specifically, when we parametrize the original SD by a Lorentzian function
\bea\label{EQ:spectral_density_old}
\Gamma_\alpha^{(0)}(\omega) = \frac{\Gamma_\alpha \delta_\alpha^2}{(\omega-\epsilon_\alpha)^2 + \delta_\alpha^2}
\eea
the TQD system parameters can be analytically evaluated and the transformed SD becomes flat
\bea\label{EQ:spectral_density_new}
\Omega_\alpha &=& \epsilon_\alpha\,,\qquad
\lambda_\alpha = \sqrt{\frac{\Gamma_\alpha \delta_\alpha}{2}}\,,\nn
\Gamma_\alpha^{(1)}(\omega) &=& 2\delta_\alpha\,.
\eea
This suggests that a Markovian treatment of the TQD system in the infinite-bias and/or high temperature regime yields the exact dynamics 
of the SQD.
We note that opposed to previous treatments of similar mappings in the literature (e.g. Refs.~\cite{elattari2000a,zedler2009a}), the mapping discussed above
can be systematically extended by mapping any reservoir into a chain of reaction coordinates.
Most important however, the discussed mapping holds also for time-dependent modifications (see also Ref.~\cite{restrepo2018a} for a bosonic periodically driven example).
In the transformed picture, instead of changing the coupling to the residual reservoirs, the feedback loop now modifies a parameter ($\lambda_\alpha$) of the TQD Hamiltonian only.
This means that in case of the discussed piecewise-constant feedback interventions we simply have 
$H_{\rm S} \to H_{\rm S}^{{\rm E}/{\rm F}}$, whereas the coupling to the residual reservoirs remains constant.
%

%%%%%%%%%%%%%%%%%%%%%%%%%%%%%%%%%%%%%%%%%%%%%%%%%%%%%%%%%%%%%%%%%%%%%%%%%%%%%%%%%%%%%%%%%%%%%%%%%%%
%%%%%%%%%%%%%%%%%%%%%%%%%%%%%%%%%%%%%%%%%%%%%%%%%%%%%%%%%%%%%%%%%%%%%%%%%%%%%%%%%%%%%%%%%%%%%%%%%%%

\subsection{TQD with projective measurements and Zeno blockade}\label{SEC:projective}

Let $P_{\rm  E} = d d^\dagger$ and $P_{\rm F} = d^\dagger d$ denote projection operators on the empty or filled central dot state, respectively.
Upon measuring outcome $\nu\in\{{\rm E},{\rm F}\}$ with probability $\trace{P_\nu \rho}$, the TQD density matrix transforms according to 
\bea
\rho^{\nu} = \frac{P_\nu \rho P_\nu}{\trace{P_\nu \rho}}\,.
\eea
Now, if for each measurement outcome the subsequent evolution $\dot\rho = {\cal L}_\nu \rho$ is conditioned on the measurement outcome $\nu$ 
(which results from switching $H_S \to H_S^\nu$), we get -- by averaging over the measurement outcomes -- immediately before the next measurement
\bea\label{EQ:dm_iteration}
\rho(t+\Delta t) = \left[e^{{\cal L}_{\rm E} \Delta t} {\cal P}_{\rm E} + e^{{\cal L}_{\rm F} \Delta t} {\cal P}_{\rm F}\right] \rho(t)\,,
\eea
where we have used ${\cal P}_\nu \rho \hat{=} P_\nu \rho P_\nu$.
We see that even in absence of feedback (${\cal L}_{\rm E}={\cal L}_{\rm F}$), the projection superoperators ${\cal P}_\nu$ may strongly affect the dynamics.
We note also that although $P_{\rm E}+P_{\rm F} = \f{1}$ in ordinary operator space, this does not hold in 
superoperator space ${\cal P}_{\rm E} + {\cal P}_{\rm F} \neq 1$, which formally reflects the fact
that quantum measurements always affect the system.
This prevents the transformation of Eq.~(\ref{EQ:dm_iteration}) into a master equation in the continuum limit ($\Delta t \to 0$).
However, we can use that ${\cal P}_\nu {\cal P}_{\nu'} = \delta_{\nu\nu'} {\cal P}_\nu$ to infer that
the projected density matrix $\hat\rho \equiv ({\cal P}_{\rm E} + {\cal P}_{\rm F}) \rho \hat{=} P_{\rm E} \rho P_{\rm E} + P_{\rm F} \rho P_{\rm F}$
obeys in the continuous measurement limit a master equation of the form
\bea
\dot{\hat{\rho}} = ({\cal P}_{\rm E} + {\cal P}_{\rm F}) \left({\cal L}_{\rm E} {\cal P}_{\rm E} + {\cal L}_{\rm F} {\cal P}_{\rm F}\right) \hat\rho \equiv {\cal L}_{\rm fb}^{\rm prj} \hat\rho\,.
\eea
In numerical investigations (not shown) we have found that this effective feedback Liouvillian 
${\cal L}_{\rm fb}^{\rm prj}$ is bistable, with different stationary
solutions corresponding to an empty and filled central dot, respectively.
In addition, the currents associated with these stationary states vanish throughout.
Thus, the usual Redfield treatment -- see App.~\ref{APP:bornmarkovme} -- will lead to a complete blockade of the current if the central dot 
is strongly and continuously measured, see also Sec.~\ref{SEC:heatflow_control}.

This can be attributed to a Zeno-type blocking of transport~\cite{elattari2000a,gurvitz2006a}, which however is not observed in actual electronic transport experiments.
This motivates us to model the effect of measurement more realistically, which naturally leads to the concept of positive operator-valued measures (POVMs)~\cite{nielsen2000,wiseman2010} or
weak measurements.
We note that the secular approximation would not imply a TQD current blockade, but it is not applicable in the continuum measurement limit.
The Zeno blockade of the current for projective measurements at high rates is also found in an independent 
investigation~\cite{engelhardt2017c} based on dynamical coarse-graining~\cite{schaller2008a}.

%%%%%%%%%%%%%%%%%%%%%%%%%%%%%%%%%%%%%%%%%%%%%%%%%%%%%%%%%%%%%%%%%%%%%%%%%%%%%%%%%%%%%%%%%%%%%%%%%%%
%%%%%%%%%%%%%%%%%%%%%%%%%%%%%%%%%%%%%%%%%%%%%%%%%%%%%%%%%%%%%%%%%%%%%%%%%%%%%%%%%%%%%%%%%%%%%%%%%%%

\subsection{TQD with weak measurements}\label{SEC:weak}

A natural way to introduce a weak measurement is via a physical interaction with a detection device such as a QPC~\cite{averin2005a,schaller2014}.
Schematically, system and detector are allowed to interact for a finite time $\Delta\tau$ (described e.g. by unitary or dissipative evolution), leading to the buildup of
system-detector correlations. 
Afterwards, a projective measurement in the detector Hilbert space (in our case, fixing the number of charges $n$ tunneled through the QPC during $\Delta\tau$) 
performs a weak measurement on the system (TQD), implementing Neumarks theorem~\cite{peres1990a}.
The feedback loop could then be closed by conditioning the subsequent evolution on the measurement outcome, as sketched in Fig.~\ref{FIG:feedback_setup_sketch} bottom panel.
To characterize the measurement properties, we will for the moment however not consider any feedback and consider the limit 
$\Delta\tau=\Delta t$ (measurement device is always on).

\begin{figure}[ht]
\includegraphics[width=0.49\textwidth,clip=true]{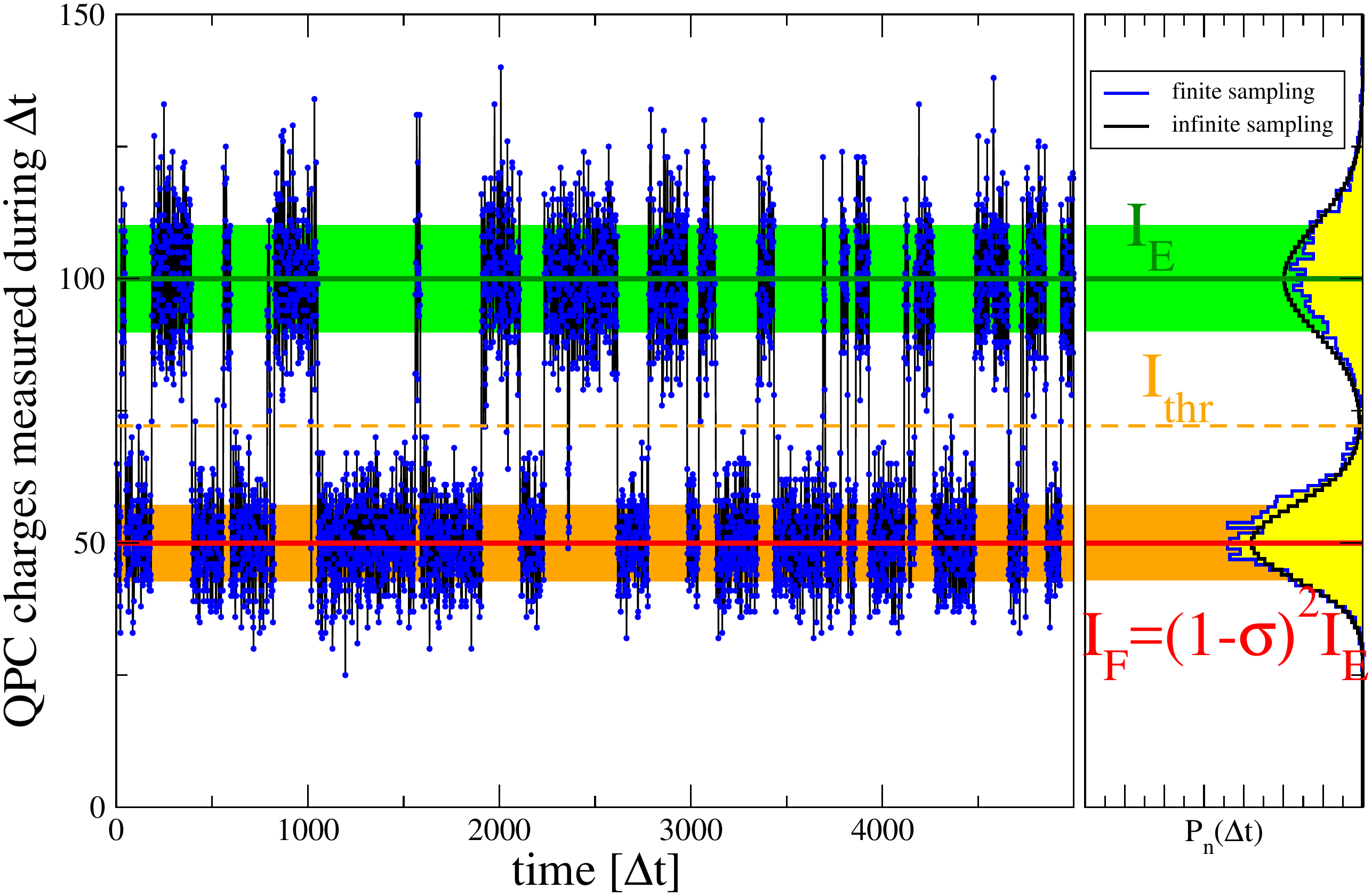}
\caption{\label{FIG:cvecqpc_measurement}
Left: Illustrative QPC trajectory described by the number of charges transferred through the QPC during time interval $\Delta\tau=\Delta t$.
Dividing the particle number by $\Delta t$ defines a time-dependent current.
The distribution of tunneled charges (right) clusters around two mean values, which correspond to $x=I_E\Delta t=\gamma\Delta t$ 
(upper green line) and $y=I_F\Delta t = \gamma(1-\sigma)^2\Delta t$ (lower red line).
The detector can discriminate between the two charge states of the central dot when the two peaks of the distribution are 
sufficiently separate. In the extreme limit, for $x\gg y \gg 1$, the projective measurement is reproduced.
The shown trajectory (an extension of Fig.~5.12 of Ref.~\cite{schaller2014}) has been generated for an SQD monitored by a weakly-transparent 
QPC, compare App.~\ref{APP:example_set_qpc}. 
Parameters: $f_L(\epsilon) = 1.0$, $f_R(\epsilon)=0.0$, $\Gamma_L\Delta t = \Gamma_R \Delta t = 0.01$, 
$\gamma \Delta t = 100$, $\gamma \Delta t (1-\sigma)^2 = 50$.
}
\end{figure}

Starting from a microscopic model for the interaction between the TQD and a QPC measurement device, we derive an effective
Lindblad generator for the TQD dynamics during the measurement, which eventually can be used to obtain the weak measurement superoperator, see Appendix~\ref{APP:measurement}.
Effectively, the weak measurement is described by a POVM, which depends on two dimensionless parameters $x$ and $y$ and
can be written as a minimally disturbing measurement~\cite{wiseman2010}, which after observing $n$ tunneled QPC electrons during measurement interval $\Delta \tau$
acts on the TQD system density matrix $\rho$ as
\bea\label{EQ:povm_measurement}
{\cal M}_n \rho &=& M_n \rho M_n^\dagger\,,\nn
M_n &=& \frac{x^{n/2}}{\sqrt{n!}} e^{-x/2} d d^\dagger + \frac{y^{n/2}}{\sqrt{n!}} e^{-y/2} d^\dagger d = M_n^\dagger\,.
\eea
The special form of our detector model lets the measurement affect the central dot only.
It is not hard to show that $\sum_n M_n^\dagger M_n = \f{1}$ although the $M_n$ operators are no projectors.
Microscopically, the $x$ and $y$ parameters are linked to the maximum QPC current $\gamma$, the reduced QPC current $\gamma(1-\sigma)^2$, and the measurement time $\Delta\tau$.
They correspond to the average particle transfer through the QPC during $\Delta t$ for an empty ($x=\gamma\Delta\tau$) or filled ($y=\gamma\Delta\tau(1-\sigma)^2$)
SQD, respectively, see Fig.~\ref{FIG:cvecqpc_measurement}.

Formally, we see directly that the action of ${\cal M}_n$ preserves the Hermiticity of any valid density matrix.
Furthermore, the positivity of $\rho$ is also preserved, which can be deduced from the observation that both trace and determinant of ${\cal M}_n \rho$
are non-negative for any valid $\rho$.
To preserve the trace, one has to renormalize afterwards, i.e., 
$\rho^{(n)} = \frac{{\cal M}_n \rho}{\trace{{\cal M}_n \rho}}$ is a valid density matrix.
The limit $x=y$ corresponds to a QPC that is insensitive to the dot occupation ($\sigma=0$), and consequently it has (after normalization) 
no effect on the TQD.
Usually, the detector is tuned to obtain information about the system, and the von-Neumann entropy of the system
\bea
S_{\rm vN}(\rho) = -\trace{\rho \ln \rho}
\eea
will decrease for most individual measurement outcomes.
However, this need not always be the case, i.e., the entropy for individual outcomes $n$ may also increase under the action of the measurement.
In particular, on average, the effect of any minimally disturbing measurement will increase the entropy~\cite{wiseman2010} $S_{\rm vN}(\sum_n {\cal M}_n \rho) \ge S_{\rm vN}(\rho)$.
For the model at hand one can confirm this by considering that on average the measurement will induce a reduction of coherences, bringing the eigenvalues of the 
reduced central dot density matrix closer together and thereby increasing the entropy.

Since the QPC statistics is in the considered limit just given by the sum of two Poissonian distributions that propagate at different speeds, 
one can define a suitable discrimination threshold (dashed line and right panel in Fig.~\ref{FIG:cvecqpc_measurement}) at the point
where both distributions coincide, i.e., where $x^{n_{\rm thr}} e^{-x} = y^{n_{\rm thr}} e^{-y}$, leading to
\bea
n_{\rm thr} = I_{\rm thr} \Delta\tau = \frac{x-y}{\ln(x)-\ln(y)}\,.
\eea
Supposing that $x\gg y$ (sensitive QPC), obtaining a value of $n$ in the measurement that is close to the two peaks will tell a lot about the 
state of the system, but when $n\approx n_{\rm thr}$, the measurement is practically useless.
To obtain a compact description with just two rates (and also to compare with the SQD treatment) 
it is natural to coarse-grain the measurement outcomes into the ones interpreted as an empty or filled dot, respectively
\bea
{\cal M}_{\rm E} \rho &\equiv& \sum_{n \ge n_{\rm thr}} {\cal M}_n\rho\\
&\hat{=}& F_{n_{\rm thr}}(x) d d^\dagger \rho d d^\dagger + F_{n_{\rm thr}}(y) d^\dagger d \rho d^\dagger d\nn 
&&+ e^{-\frac{(\sqrt{x}-\sqrt{y})^2}{2}} F_{n_{\rm thr}}(\sqrt{xy})\left(d^\dagger d \rho d d^\dagger + d d^\dagger \rho d^\dagger d\right)\,,\nn
{\cal M}_{\rm F}\rho &\equiv& \sum_{n < n_{\rm thr}} {\cal M}_n\rho\nn
&\hat{=}& [1-F_{n_{\rm thr}}(x)] d d^\dagger \rho d d^\dagger + [1-F_{n_{\rm thr}}(y)] d^\dagger d \rho d^\dagger d\nn
&&+ e^{-\frac{(\sqrt{x}-\sqrt{y})^2}{2}} [1-F_{n_{\rm thr}}(\sqrt{xy})]\left(d^\dagger d \rho d d^\dagger + {\rm h.c.}\right)\,,\nonumber %d d^\dagger \rho d^\dagger d\right)\,,
\eea 
where $\rho$ still denotes the TQD density matrix and $F_n(x) \equiv \Gamma(n,x)/\Gamma(n)$ with $\Gamma(n,x)$ 
denoting the incomplete Gamma function and $\Gamma(n) = (n-1)!$ the ordinary Gamma function.
The $F_n(x)$ functions behave similar to Fermi functions, such that when $x \gg y \gg 1$, the measurement superoperators approach the projective 
limit ${\cal M}_{{\rm E}/{\rm F}} \to {\cal P}_{{\rm E }/{\rm F}}$.
We note that, as for the projective case, these superoperators do not add up to the identity.
%
%Rather, it follows from Fourier transform properties that ${\cal M}_{\rm E} + {\cal M}_{\rm F} = e^{{\cal L}_{\rm dt}(0) \Delta\tau}$, see App.~\ref{APP:measurement}.
%
Furthermore, we also note that ${\cal M}_{\rm E} {\cal M}_{\rm F} = {\cal M}_{\rm F} {\cal M}_{\rm E}$.

%%%%%%%%%%%%%%%%%%%%%%%%%%%%%%%%%%%%%%%%%%%%%%%%%%%%%%%%%%%%%%%%%%%%%%%%%%%%%%%%%%%%%%%%%%%%%%%%%%%
%%%%%%%%%%%%%%%%%%%%%%%%%%%%%%%%%%%%%%%%%%%%%%%%%%%%%%%%%%%%%%%%%%%%%%%%%%%%%%%%%%%%%%%%%%%%%%%%%%%

\subsection{TQD with feedback}\label{SEC:feedback}

By conditioning the subsequent evolution on the measurement outcome, we close the feedback loop.
We denote the implicit dependence of the dissipators on $\Gamma_\alpha^{\rm E/F}$ by ${\cal L} \to {\cal L}_{\rm E/F}$, which yields for $\Delta t > \Delta\tau$
the feedback iteration equation (recall that the measurement duration $\Delta\tau$ is implicit in the ${\cal M}_{{\rm E }/{\rm F}}$)
\bea
\rho(t+\Delta t) &=& \left(e^{{\cal L}_{\rm E} (\Delta t-\Delta\tau)}{\cal M}_{\rm E} + e^{{\cal L}_{\rm F} (\Delta t-\Delta\tau)}{\cal M}_{\rm F}\right)\rho(t)\nn
&=& {\cal P}(\Delta t) \rho(t)\,.
\eea
For finite $\Delta\tau \le \Delta t$ this can be solved for a stroboscopic stationary state $\bar\rho^{\Delta t,\Delta\tau} = {\cal P}(\Delta t) \bar\rho^{\Delta t,\Delta\tau}$.
We note that when ${\cal L}_{{\rm E }/{\rm F}}$ are of Lindblad form, the above propagator ${\cal P}(\Delta t)$ will preserve all density matrix properties, since it is derived 
from an average over all conditional evolutions (which separately preserve the density matrix properties).
This property must be preserved in the limit of small $\Delta t$.
We therefore consider the limit $\Delta\tau \ll \Delta t \to 0$, keeping however $x$ and $y$ finite to preserve the measurement effects~\cite{caves1987a}.
This implies that the QPC coupling $\gamma$ must be large, justifying a posteriori the singular coupling limit used in its derivation in App.~\ref{APP:measurement}.
Formally, we therefore only set the explicit dependence on $\Delta \tau$ to zero and then expand for small $\Delta t$ to obtain
\bea
\rho(t+\Delta t) = \left[{\cal M}_{\rm E} + {\cal M}_{\rm F} +\Delta t \left({\cal L}_{\rm E} {\cal M}_{\rm E} + {\cal L}_{\rm F} {\cal M}_{\rm F}\right)\right]\rho(t)\,.\nn
\eea 
Subtracting $\rho(t)$ on both sides and dividing by $\Delta t$ we get
an effective feedback master equation, described by the feedback dissipator
\bea\label{EQ:liouville_feedback}
{\cal L}_{\rm fb} = \frac{{\cal M}_{\rm E}+{\cal M}_{\rm F}-\f{1}}{\Delta t} +  {\cal L}_{\rm E} {\cal M}_{\rm E} + {\cal L}_{\rm F} {\cal M}_{\rm F}\,.
\eea
The corresponding stationary state will be defined by ${\cal L}_{\rm fb} \bar\rho=\f{0}$.
The first dissipator defines an effective measurement dissipator (compare App.~\ref{APP:measurement})
\bea\label{EQ:dissipator_measurement}
{\cal L}_{\rm ms}\rho &\equiv& \frac{{\cal M}_{\rm E}+{\cal M}_{\rm F}-\f{1}}{\Delta t}\rho\nn
&\hat{=}& \bar\Gamma \left[d^\dagger d \rho d^\dagger d - \frac{1}{2}\left\{d^\dagger d, \rho\right\}\right]\,,\qquad
\eea
which is of Lindblad form and appears formally divergent as $\Delta t\to 0$.
However, in the discussed regime $\Delta\tau\ll \Delta t$ we stress that the measurement dissipator remains finite.
Effectively, it is determined by the parameter 
$\bar\Gamma=\frac{2}{\Delta t} \left(1-e^{-\sigma^2\gamma\Delta\tau/2}\right)>0$ describing the dephasing due to the QPC measurement~\cite{averin2005a,gurvitz2006a}.
If we would have directly expanded the projective measurement iteration~(\ref{EQ:dm_iteration}), the corresponding measurement dissipator would indeed diverge. 
The action of (\ref{EQ:dissipator_measurement}) on the central dot density matrix appears trivial, 
since it deletes coherences which cannot be created anyway and leaves the diagonal elements of the
central dot untouched.
However, for the simulation of the full TQD density matrix we have found that the inclusion of this term with sufficiently small $\Delta t$ 
is necessary to preserve the density matrix properties of the TQD system:
Even when ${\cal L}_{\rm E}$ and ${\cal L}_{\rm F}$ are chosen of Lindblad form and ${\cal M}_{{\rm E }/{\rm F}}$ separately preserve positivity and Hermiticity
and ${\cal M}_{\rm E}+{\cal M}_{\rm F}$ corresponds to a Lindblad exponential, the action of ${\cal L}_{\rm E} {\cal M}_{\rm E} + {\cal L}_{\rm F} {\cal M}_{\rm F}$ alone will in general not
preserve the density matrix properties.
These can only be recovered by adding a very strong (strongness of the superoperator does not imply a projective measurement) measurement dissipator.
%

%%%%%%%%%%%%%%%%%%%%%%%%%%%%%%%%%%%%%%%%%%%%%%%%%%%%%%%%%%%%%%%%%%%%%%%%%%%%%%%%%%%%%%%%%%%%%%%%%%%
%%%%%%%%%%%%%%%%%%%%%%%%%%%%%%%%%%%%%%%%%%%%%%%%%%%%%%%%%%%%%%%%%%%%%%%%%%%%%%%%%%%%%%%%%%%%%%%%%%%
%%%%%%%%%%%%%%%%%%%%%%%%%%%%%%%%%%%%%%%%%%%%%%%%%%%%%%%%%%%%%%%%%%%%%%%%%%%%%%%%%%%%%%%%%%%%%%%%%%%
%%%%%%%%%%%%%%%%%%%%%%%%%%%%%%%%%%%%%%%%%%%%%%%%%%%%%%%%%%%%%%%%%%%%%%%%%%%%%%%%%%%%%%%%%%%%%%%%%%%
\section{Thermodynamics}\label{SEC:thermodynamics}

In this section, we will define sensible expressions for the heat exchange during the control and measurement
phases (Sec.~\ref{SEC:heatflow_control} and Sec.~\ref{SEC:heat_measurement}, respectively), and for the work 
required to switch the Hamiltonian between measurement and control (Sec.~\ref{SEC:switching_work}).
These can be used to show that our system obeys the first law (Sec.~\ref{SEC:first_law}) and the second law (Sec.~\ref{SEC:second_law}) of thermodynamics,
which bound the performance of the demon.

\subsection{Heat flow during conditional evolution}\label{SEC:heatflow_control}

Since by construction ${\cal L}_{\rm E} = {\cal L}_{\rm E}^{({\rm L})} + {\cal L}_{\rm E}^{({\rm R})}$ and
${\cal L}_{\rm F} = {\cal L}_{\rm F}^{({\rm L})} + {\cal L}_{\rm F}^{({\rm R})}$, the new dissipator 
${\cal L}_{\rm fb} ={\cal L}_{\rm ms} + {\cal L}_{\rm E} {\cal M}_{\rm E} + {\cal L}_{\rm F} {\cal M}_{\rm F} 
= {\cal L}_{\rm ms} + {\cal L}_{\rm fb}^{({\rm L})} + {\cal L}_{\rm fb}^{({\rm R})}$
is still additive in the reservoirs, where 
${\cal L}_{\rm fb}^{(\alpha)} = {\cal L}_{\rm E}^{(\alpha)} {\cal M}_{\rm E} +  {\cal L}_{\rm F}^{(\alpha)} {\cal M}_{\rm F}$. 
Therefore, a phenomenologic way to define the stationary matter current entering the TQD from reservoir $\alpha$
proceeds via considering $\frac{d}{dt} \expval{N_S} = \trace{(d_{\rm L}^\dagger d_{\rm L} + d^\dagger d + d_{\rm R}^\dagger d_{\rm R})\dot\rho}$, which would 
at steady state result in 
\bea
I_{\rm M}^{(\alpha)} = \trace{(d_{\rm L}^\dagger d_{\rm L} + d^\dagger d + d_{\rm R}^\dagger d_{\rm R}) ({\cal L}_{\rm fb}^{(\alpha)}\bar\rho)}\,,
\eea
where $\bar\rho$ denotes the stationary state of the full feedback Liouvillian~(\ref{EQ:liouville_feedback}) and $\alpha\in\{{\rm L},{\rm R}\}$.
First, we note that the matter current at steady state is conserved $I_{\rm M}^{({\rm L})}+I_{\rm M}^{({\rm R})}=0$, 
since the feedback operations do not inject particles into the TQD system.
We further note that the currents depend on $\delta_\alpha$, but also implicitly on $\Gamma_\alpha^{{\rm E }/{\rm F}}$ due to the feedback.
Also, doubling $\Gamma_{\rm L}$ and $\Gamma_{\rm R}$ will not necessarily double the current, since these parameters enter the TQD Hamiltonian.
An alternative definition of the steady-state matter current would be to look at the time derivative of the occupation of the central dot,
which decomposes into a left- and right-flowing contribution.
Due to the special structure of the Born-Markov master equation -- see App.~\ref{APP:bornmarkovme} -- 
these are fully identical with the Heisenberg equations of motion for the central dot, leading to the definition
$I_{\rm M}^{(\alpha)} = -\ii \sqrt{\frac{\Gamma_\alpha \delta}{2}} \trace{\left[d^\dagger d_\alpha - d_\alpha^\dagger d\right] \bar\rho}$.
From this form, we see more directly that a projected density matrix (e.g. $\bar\rho \to d^\dagger d \rho d^\dagger d$) 
would lead to a vanishing matter current for any density matrix $\rho$ (e.g. via
$\trace{\left[d^\dagger d_\alpha - d_\alpha^\dagger d\right] d^\dagger d \rho d^\dagger d} = \trace{d^\dagger d \left[d^\dagger d_\alpha - d_\alpha^\dagger d\right] d^\dagger d\rho}=0$, 
using that $d^2 = \f{0} = (d^\dagger)^2$).

Finally, the matter currents entering the TQD can also be defined microscopically using the counting field formalism~\cite{esposito2009a}.
We have found these three definitions to be equivalent at steady state within the Born-Markov (non-secular) description.

The energy currents entering the system from the left and right reservoir would be sensibly defined in a similar way.
We note that this treatment neglects the interaction energy between TQD and its reservoirs, but keeps the interaction energy between the 
central dot and its reservoirs.
However, in addition the system Hamiltonian now depends on the control operations, eventually leading to 
\bea\label{EQ:elrrate}
I_{\rm E}^{(\alpha)} = \trace{H_{\rm S}^{\rm E} ({\cal L}_{\rm E}^{(\alpha)} {\cal M}_{\rm E} \bar\rho)+H_{\rm S}^{\rm F} ({\cal L}_{\rm F}^{(\alpha)} {\cal M}_{\rm F} \bar\rho)}\,.
\eea
In presence of feedback (${\cal L}_{\rm E}^{(\alpha)} \neq {\cal L}_{\rm F}^{(\alpha)}$ and $H_{\rm S}^{\rm E} \neq H_{\rm S}^{\rm F}$), the energy currents are not necessarily conserved, 
since the feedback loop may inject energy into the system, both during measurement and switching.
Again, we find from microscopic considerations based on counting fields the same definitions for the energy exchanged with the
reservoirs.
Together, the energy and matter currents enter the heat current from the corresponding reservoirs
$\dot{Q}^{(\alpha)} = I_{\rm E}^{(\alpha)} - \mu_\alpha I_{\rm M}^{(\alpha)}$.

%%%%%%%%%%%%%%%%%%%%%%%%%%%%%%%%%%%%%%%%%%%%%%%%%%%%%%%%%%%%%%%%%%%%%%%%%%%%%%%%%%%%%%%%%%%%%%%%%%%
%%%%%%%%%%%%%%%%%%%%%%%%%%%%%%%%%%%%%%%%%%%%%%%%%%%%%%%%%%%%%%%%%%%%%%%%%%%%%%%%%%%%%%%%%%%%%%%%%%%

\subsection{Heat during measurement}\label{SEC:heat_measurement}

If we do not change the TQD Hamiltonian $H_{\rm S}$ during the measurement, the average energy injected into the TQD during $\Delta \tau$ is (in App.~\ref{APP:measurement} we detail 
${\cal M}_E + {\cal M}_F$ using a microscopic model)
\bea
\Delta E_{\rm ms} &=& \trace{H_{\rm S} \left({\cal M}_{\rm E} + {\cal M}_{\rm F} - \f{1}\right) \rho}\nn
&=& \Delta t \trace{H_{\rm S} {\cal L}_{\rm ms} \rho}\,,
\eea
which suggests to define the energy current due to the measurement in the conventional way
\bea\label{EQ:msrate}
I_{\rm E}^{\rm ms} = \trace{H_{\rm S} {\cal L}_{\rm ms} \rho}\,.
\eea
Making the form of ${\cal L}_{\rm ms}$ explicit, we see that when 
$[d^\dagger d,H_{\rm S}]\to 0$ (this holds approximately in the weak coupling limit between central dot and its reservoirs), 
the measurement will on average not inject any energy.
Furthermore, the measurement-associated energy current also vanishes when the measurement is insensitive ($\sigma=0$).
For simplicity, we use as a natural choice the average of the two Hamiltonians $H_{\rm S} = \frac{1}{2} \left(H_{\rm S}^{\rm E} + H_{\rm S}^{\rm F}\right)$ in the numerical results of this
paper.

We finally note that a special Hamiltonian acting during the measurement implies that switching work is applied twice to the TQD, at the beginning and at the end of the measurement process, see below.
An alternative scheme would be to leave the previous Hamiltonian acting during the measurement, which however would complicate the discussion.

%%%%%%%%%%%%%%%%%%%%%%%%%%%%%%%%%%%%%%%%%%%%%%%%%%%%%%%%%%%%%%%%%%%%%%%%%%%%%%%%%%%%%%%%%%%%%%%%%%%
%%%%%%%%%%%%%%%%%%%%%%%%%%%%%%%%%%%%%%%%%%%%%%%%%%%%%%%%%%%%%%%%%%%%%%%%%%%%%%%%%%%%%%%%%%%%%%%%%%%

\subsection{Switching work}\label{SEC:switching_work}

To run the feedback loop, work has to be performed on the system, both when initializing the measurement (switching the Hamiltonian from $H_{\rm S}^{{\rm E }/{\rm F}}$ to $H_{\rm S}$) 
and right after the measurement (switching back from $H_{\rm S}$ to $H_{\rm S}^{{\rm E }/{\rm F}}$).
The Hamiltonian $H_{\rm S}$ chosen implicitly determines the rates $\Gamma_\alpha$ during the measurement via Eq.~(\ref{EQ:ham_mapped}),
compare also Fig.~\ref{FIG:feedback_setup_sketch}.
On average, this will imply for the switching work during a feedback cycle
\bea
W_{\rm sw} &=& 
\trace{(H_{\rm S} - H_{\rm S}^{\rm E}) e^{{\cal L}_{\rm E} \Delta t} {\cal M}_{\rm E} \bar\rho}\nn
&&+ \trace{(H_{\rm S} - H_{\rm S}^{\rm F}) e^{{\cal L}_{\rm F} \Delta t} {\cal M}_{\rm F} \bar\rho}\nn
&&+\trace{(H_{\rm S}^{\rm E}-H_{\rm S}) {\cal M}_{\rm E} \bar\rho}\nn
&&+ \trace{(H_{\rm S}^{\rm F}-H_{\rm S}) {\cal M}_{\rm F} \bar\rho}\,.
\eea
Now, upon expanding for small $\Delta t$ we see that the leading order in $\Delta t$ is linear, and we get
for the $\dot{W}_{\rm sw} = \frac{W_{\rm sw}}{\Delta t}$ the expression
\bea\label{EQ:workrate}
\dot{W}_{\rm sw} &=& \trace{(H_{\rm S}-H_{\rm S}^{\rm E}) ({\cal L}_{\rm E} {\cal M}_{\rm E} \bar\rho)}\nn
&& + \trace{(H_{\rm S}-H_{\rm S}^{\rm F}) ({\cal L}_{\rm F} {\cal M}_{\rm F} \bar\rho)}\,.
\eea
We see that for a constant TQD Hamiltonian throughout, this expression must vanish.
It can be further simplified by adopting the choice 
$H_{\rm S} = 1/2(H_{\rm S}^{\rm E}+H_{\rm S}^{\rm F})$ (which we will use in our numerical simulations), 
where the expression for the work rate becomes
$\dot{W}_{\rm sw} = \frac{1}{2} \trace{(H_{\rm S}^{\rm F}-H_{\rm S}^{\rm E}) ({\cal L}_{\rm E} {\cal M}_{\rm E} - {\cal L_{\rm F}} {\cal M}_{\rm F})\bar\rho}$.
In particular, with $\delta_L=\delta_R$ this means for the rates during measurement 
$\Gamma_\alpha = \frac{1}{4}\left(\sqrt{\Gamma_\alpha^E} + \sqrt{\Gamma_\alpha^F}\right)^2$.

%%%%%%%%%%%%%%%%%%%%%%%%%%%%%%%%%%%%%%%%%%%%%%%%%%%%%%%%%%%%%%%%%%%%%%%%%%%%%%%%%%%%%%%%%%%%%%%%%%%
%%%%%%%%%%%%%%%%%%%%%%%%%%%%%%%%%%%%%%%%%%%%%%%%%%%%%%%%%%%%%%%%%%%%%%%%%%%%%%%%%%%%%%%%%%%%%%%%%%%
\subsection{First law}\label{SEC:first_law}

By adding the contributions (\ref{EQ:elrrate}), (\ref{EQ:msrate}), and (\ref{EQ:workrate}) we see analytically that at steady state the energy is conserved
\bea
0 &=& I_{\rm E}^{({\rm L})} + I_{\rm E}^{({\rm R})} + I_{\rm E}^{\rm ms} + \dot{W}_{\rm sw}\nn
&=& \trace{H_{\rm S} \left({\cal L}_{\rm ms} + {\cal L}_{\rm E} {\cal M}_{\rm E} + {\cal L}_{\rm F} {\cal M}_{\rm F}\right) \bar\rho)}\,,
\eea
where we have used that ${\cal L}_{{\rm E }/{\rm F}} = {\cal L}_{{\rm E }/{\rm F}}^{({\rm L})} + {\cal L}_{{\rm E }/{\rm F}}^{({\rm R})}$.
This expression vanishes as $\bar\rho$ denotes the stationary state of the corresponding feedback Liouvillian.
In our considerations we will only investigate the total energy flow due to the feedback loop
\bea\label{EQ:energy_conservation}
I_{\rm E}^{\rm fb} = I_{\rm E}^{\rm ms} + \dot{W}_{\rm sw} = - \left(I_{\rm E}^{({\rm L})} + I_{\rm E}^{({\rm R})}\right)\,,
\eea
which is manifest already by a mismatch of left and right energy currents.
Considering the generation of electric power $P_{\rm el} = -(\mu_{\rm L}-\mu_{\rm R}) I_{\rm M}^{({\rm L})}$ as the main demon objective, it appears natural to define the gain
as the ratio of output power vs. the energy required to run the feedback loop
\bea\label{EQ:gain}
G \equiv \frac{P_{\rm el}}{\dot{W}_{\rm sw} + I_{\rm E}^{\rm ms}} = \frac{V I_{\rm M}^{({\rm L})}}{I_{\rm E}^{({\rm L})} + I_{\rm E}^{({\rm R})}}\,.
\eea
Since it is also information that is needed to run the feedback loop, the gain is not bounded by one.
Furthermore, since it counts only electric output power, this measure vanishes at equilibrium ($V=0$) although the demon feedback loop may produce a finite current.

%%%%%%%%%%%%%%%%%%%%%%%%%%%%%%%%%%%%%%%%%%%%%%%%%%%%%%%%%%%%%%%%%%%%%%%%%%%%%%%%%%%%%%%%%%%%%%%%%%%
%%%%%%%%%%%%%%%%%%%%%%%%%%%%%%%%%%%%%%%%%%%%%%%%%%%%%%%%%%%%%%%%%%%%%%%%%%%%%%%%%%%%%%%%%%%%%%%%%%%
\subsection{Second law}\label{SEC:second_law}

In our considerations, we are operating the QPC in a unidirectional transport regime, whose associated entropy production rate~\cite{esposito2009a}
$\dot{S}_\ii =  \beta V_{\rm QPC} I_{\rm QPC}$ diverges.
With all other quantities remaining finite, the {\em global} entropy production rate is therefore always positive by construction.
However, we may attempt to write a balance equation for the {\em local} entropy of the TQD system.
In doing so, we first note that the construction of the feedback loop enables us to consider the change of the von-Neumann entropy along particular trajectories
belonging to different measurement results.

We will discuss only trajectories that start with the stationary state of the feedback loop.
Without the coarse-graining of the measurement outcomes, this is defined 
by $\bar\rho = \sum_n e^{{\cal L}_n \Delta t} {\cal M}_n \bar\rho$ 
(for simplicity of notation we have put here $\Delta\tau\to 0$).
Then, the entropy at the end of the feedback loop
will for an initial measurement outcome $n$ be given by
\bea
S^{(n)}(\Delta t) &=& S_{\rm vN}\left(\frac{e^{{\cal L}_n \Delta t} {\cal M}_n \bar\rho}{p_n}\right)\nn
&=& \Delta S^{(n)}_{\rm ct} + \Delta S^{(n)}_{\rm ms} + S_{\rm vN}(\bar\rho)\,.
\eea
Here, the probability for this outcome is given by $p_n = \trace{{\cal M}_n \bar\rho}$, $\Delta S^{(n)}_{\rm ct}$ denotes the system entropy change during control, and $\Delta S_{\rm ms}^{(n)}$
the system entropy change during measurement, all conditioned on the measurement outcome $n$.

For a Lindblad-type conditional control with thermal reservoirs, we can split the entropy change of the system 
into a non-negative irreversible part $\Delta_\ii S^{(n)}\ge 0$ and an exchange part~\cite{spohn1978a,esposito2010b}
\bea
\Delta S^{(n)}_{\rm ct} &=& \Delta_\ii S^{(n)} + \Delta_{\rm e} S^{(n)}\nn
&=& \Delta_\ii S^{(n)} + \sum_\alpha \beta_\alpha \left[\Delta E_\alpha^{(n)} - \mu_\alpha \Delta N_\alpha^{(n)}\right]\,,
\eea
where the latter part is related to the heat flows entering the system.
Inserting and solving for the exchange entropy and measurement contribution yields
\bea
-\sum_\alpha \beta_\alpha \left[\Delta E_\alpha^{(n)} - \mu_\alpha \Delta N_\alpha^{(n)}\right] - \Delta S^{(n)}_{\rm ms}=\nn 
\Delta_\ii S^{(n)} + \left[S_{\rm vN}(\bar\rho)-S_{\rm vN}\left(\frac{e^{{\cal L}_n \Delta t} {\cal M}_n \bar\rho}{p_n}\right)\right]\,.
\eea
If we average over all measurement outcomes, the last term on the r.h.s. corresponds to the mutual information between system 
and detector that is discarded, see App.~\ref{APP:repeated_interactions}.
By averaging over this expression, we see by invoking that~\cite{bergou2013} 
\bea\label{EQ:ent_inequality}
\sum_n p_n S_{\rm vN}\left(\frac{e^{{\cal L}_n \Delta t} {\cal M}_n \bar\rho}{p_n}\right) &\le& 
S_{\rm vN}\left(\sum_n p_n \frac{e^{{\cal L}_n \Delta t} {\cal M}_n \bar\rho}{p_n}\right)\nn
&=& S_{\rm vN}(\bar\rho)
\eea
and $\Delta_\ii S^{(n)} \ge 0$
on average, we must have
\bea
\sum_n p_n \left[-\sum_\alpha \beta_\alpha \left[\Delta E_\alpha^{(n)} - \mu_\alpha \Delta N_\alpha^{(n)}\right] - \Delta S^{(n)}_{\rm ms}\right] \ge 0\,.\qquad
\eea
Dividing by $\Delta t$ and considering $\Delta t \to 0$, the first terms just become the energy and matter currents, leading to 
\bea\label{EQ:second_law1}
-\sum_\alpha \beta_\alpha \left[I_{\rm E}^{(\alpha)} - \mu_\alpha I_{\rm M}^{(\alpha)}\right] - d_t S_{\rm ms} \ge 0\,,
\eea
which denotes a version of the second law for the continuum weak measurement limit~\cite{jacobs2009a}.
It can be used to bound the energetic performance of the device by the information gained by measurement.
For our setup where $\beta_{\rm L}=\beta_{\rm R}=\beta$ this yields -- using Eq.~(\ref{EQ:energy_conservation}) -- our version
of the second law
\bea\label{EQ:second_law2}
\beta(I_{\rm E}^{\rm ms} + \dot{W}_{\rm sw} - P_{\rm el}) - d_t S_{\rm ms} \ge 0\,,
\eea
which can be used to bound e.g. the gain. 
In particular, to have a gain $G>1$ (information-driven regime) when $I_{\rm E}^{\rm ms} + \dot{W}_{\rm sw} > 0$, 
it is necessary that $d_t S_{\rm ms} < 0$, i.e., that the measurement on average reduces the system entropy.
Technically, we note that the average of the measurement entropy change is given by 
\bea\label{EQ:average_measurement_entropy}
\Delta S_{\rm ms} &=& \sum_n p_n \left[\trace{\bar\rho\ln\bar\rho}-\trace{\frac{{\cal M}_n \bar\rho}{p_n} \ln \frac{{\cal M}_n \bar\rho}{p_n}} \right]\,.\qquad
\eea
In the regime of positive electric power, we can also define
an efficiency for the conversion of both information and feedback energy into electric power via (compare also Ref.~\cite{brandner2015a})
\bea\label{EQ:efficiency}
\eta = \frac{\beta P_{\rm el}}{\beta(I_{\rm E}^{\rm ms} + \dot{W}_{\rm sw}) - \frac{\Delta S_{\rm ms}}{\Delta t}}\,,
\eea
which is bounded by one by the second law~(\ref{EQ:second_law2}).
However, we stress that also other regimes are conceivable, for example generating both electric power $P_{\rm el}>0$ 
and simultaneously extracting work $\dot{W}_{\rm sw}<0$~\cite{engelhardt2017c}, which would motivate other definitions of efficiency.

Finally, we remark that the same second law can be derived when the entropic contribution of the abstract detector 
(and thus, the mutual information between system and detector) is
explicitly taken into account, see App.~\ref{APP:repeated_interactions}, which is similar to the framework of
repeated interactions~\cite{strasberg2017a}.

%%%%%%%%%%%%%%%%%%%%%%%%%%%%%%%%%%%%%%%%%%%%%%%%%%%%%%%%%%%%%%%%%%%%%%%%%%%%%%%%%%%%%%%%%%%%%%%%%%%
%%%%%%%%%%%%%%%%%%%%%%%%%%%%%%%%%%%%%%%%%%%%%%%%%%%%%%%%%%%%%%%%%%%%%%%%%%%%%%%%%%%%%%%%%%%%%%%%%%%
%%%%%%%%%%%%%%%%%%%%%%%%%%%%%%%%%%%%%%%%%%%%%%%%%%%%%%%%%%%%%%%%%%%%%%%%%%%%%%%%%%%%%%%%%%%%%%%%%%%
%%%%%%%%%%%%%%%%%%%%%%%%%%%%%%%%%%%%%%%%%%%%%%%%%%%%%%%%%%%%%%%%%%%%%%%%%%%%%%%%%%%%%%%%%%%%%%%%%%%

\section{Numerical results}\label{SEC:results}

We will first investigate the weak-coupling regime, where one would expect that the TQD treatment is
equivalent to the SQD treatment in Ref.~\cite{schaller2011b} -- as far as the current through the system is concerned.
However, our extended description now allows to quantify the injection of energy into the TQD system by measurement and control steps,
which will in general not vanish.
We will demonstrate that in the weak-coupling regime this is indeed negligibly small in comparison to the generated electric power,
such that the device indeed implements a Maxwell demon feedback loop in the weak-coupling regime.
Next, we will investigate how these relations change beyond weak-coupling.

%%%%%%%%%%%%%%%%%%%%%%%%%%%%%%%%%%%%%%%%%%%%%%%%%%%%%%%%%%%%%%%%%%%%%%%%%%%%%%%%%%%%%%%%%%%%%%%%%%%
%%%%%%%%%%%%%%%%%%%%%%%%%%%%%%%%%%%%%%%%%%%%%%%%%%%%%%%%%%%%%%%%%%%%%%%%%%%%%%%%%%%%%%%%%%%%%%%%%%%

\subsection{Weak-coupling regime}

\begin{figure*}[ht]
\includegraphics[width=0.8\textwidth,clip=true]{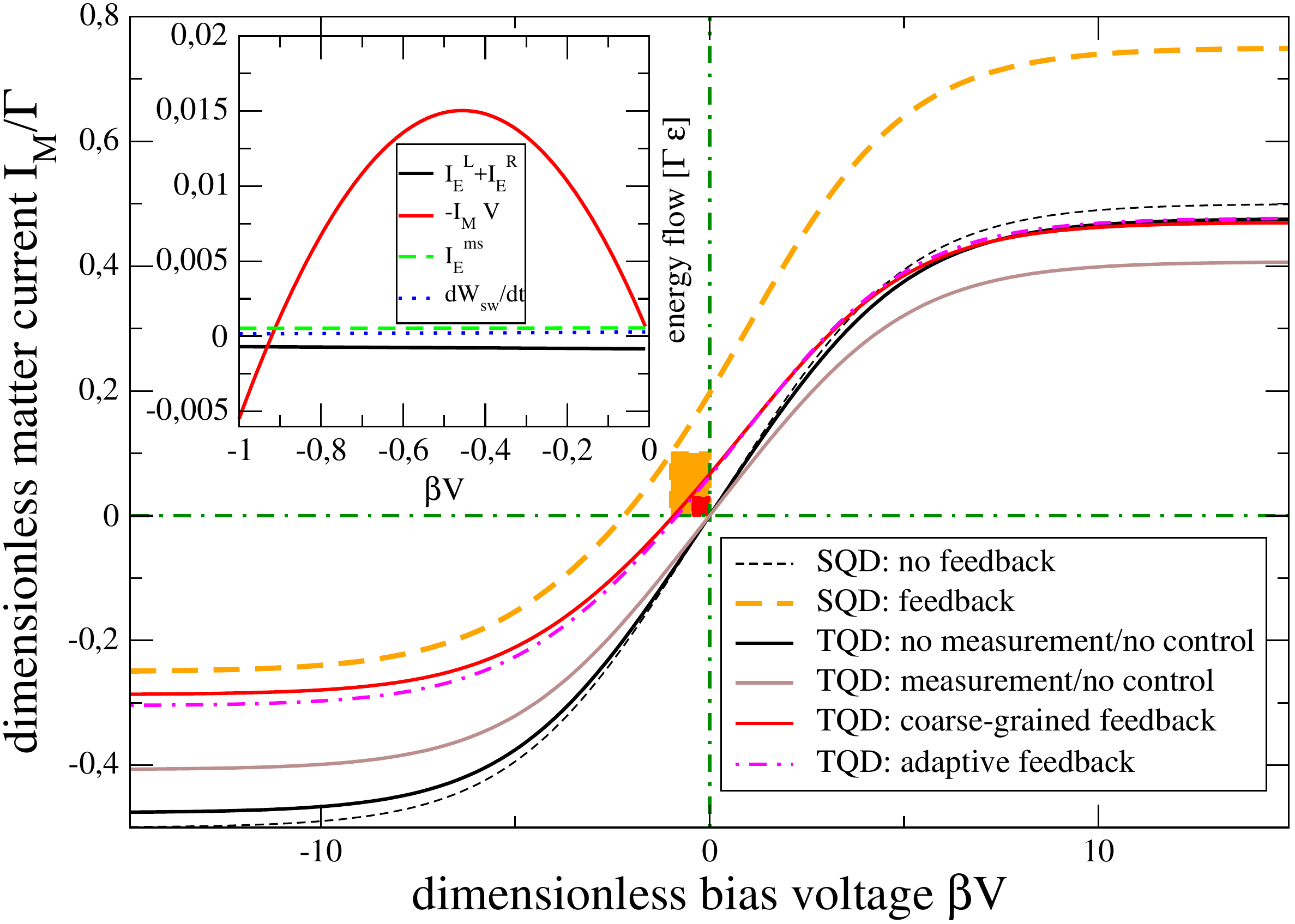}
\caption{\label{FIG:feedback_effect}
Main: Current through the TQD/SQD vs. bias voltage.
Dashed curves correspond to the SQD benchmark either in absence (thin black) or in presence of feedback (bold orange), where
finite electric power is generated (area of large rectangle).
In absence of both measurements and control actions ($x=y$ and $H_{\rm S}^{\rm E}=H_{\rm S}^{\rm F}$), the TQD treatment (solid black) 
follows the SQD treatment closely.
This changes already when measurement is active but no control is applied (solid brown).
When the feedback loop is closed (solid red), the current no longer vanishes at the origin, and there is a regime where electric power is produced, 
albeit reduced as compared to the SQD (red small rectangle).
Additionally adapting the feedback to the actual measurement outcome has little impact (dash-dotted magenta, see Sec.~\ref{SEC:coarsegraining}).
Inset: Curves for the coarse-grained feedback demonstrate that in the demon regime the generated power (solid red)
is significantly larger than other contributions such as measurement energy (dashed green), 
switching work rate (dotted blue), and sum of left and right energy currents (solid black).
Other parameters:
$\beta\epsilon=\beta\epsilon_{\rm L}=\beta\epsilon_{\rm R}=1$, $\beta\Gamma_{\rm L}^{\rm E}=\beta\Gamma_{\rm R}^{\rm F} = 0.015$, $\beta\Gamma_{\rm L}^{\rm F} = \beta \Gamma_{\rm R}^{\rm E} = 0.005$, $x=10$, $y=3$, 
$\Gamma=(\Gamma_\alpha^{\rm E} + \Gamma_\alpha^{\rm F})/2$, $\Gamma\Delta t = 1$, $\mu_{\rm L}=-\mu_{\rm R}=V/2$, 
and $\beta\delta_{\rm L}=\beta\delta_{\rm R}=0.1$.
}
\end{figure*}

We first benchmark our TQD treatment in absence of measurement ($x=y$) and also in absence of control
$H_{\rm S}^{\rm E}=H_{\rm S}^{\rm F}$ to yield similar results as the SQD treatment.
Indeed, the black curves in Fig.~\ref{FIG:feedback_effect} demonstrate close agreement of the TQD and SQD 
treatments in the weak-coupling-limit in absence of any measurement and control.
Then, we compare the SQD treatment in presence of feedback control with the TQD treatment, first only in presence
of measurements ($x>y$) but absence of control ($H_{\rm S}^{\rm E}=H_{\rm S}^{\rm F}$).
This already suppresses the current due to the partial projection of spatial superpositions (solid brown), but does not break the detailed balance
relations and therefore does not produce electric power.
Finally, when control is applied to the TQD (solid red), a similar situation as with the SQD in presence of feedback (dashed orange)
arises.
Although the electric power is significantly reduced (filled rectangle area vs. hollow rectangle area), the inset demonstrating the TQD energy
flows defined in Eqns.~(\ref{EQ:elrrate}),~(\ref{EQ:msrate}), and~(\ref{EQ:workrate}) show that these contributions are negligible in comparison to 
the electric power (solid red curve in the inset), which justifies to call this parameter regime information-dominated.
The dash-dotted magenta curve describes adaptive feedback, explained in Sec.~\ref{SEC:coarsegraining}.

In a nutshell, we obtain that the more realistic TQD treatment with weak measurements and coherent control
supports a Maxwell-demon mode in the weak-coupling regime, but with a significant reduction in electric output power.
To compensate for this, one can explore the strong-coupling regime, see below.

%%%%%%%%%%%%%%%%%%%%%%%%%%%%%%%%%%%%%%%%%%%%%%%%%%%%%%%%%%%%%%%%%%%%%%%%%%%%%%%%%%%%%%%%%%%%%%%%%%%
%%%%%%%%%%%%%%%%%%%%%%%%%%%%%%%%%%%%%%%%%%%%%%%%%%%%%%%%%%%%%%%%%%%%%%%%%%%%%%%%%%%%%%%%%%%%%%%%%%%

\subsection{Towards strong coupling}

\begin{figure*}[ht]
\includegraphics[width=0.8\textwidth,clip=true]{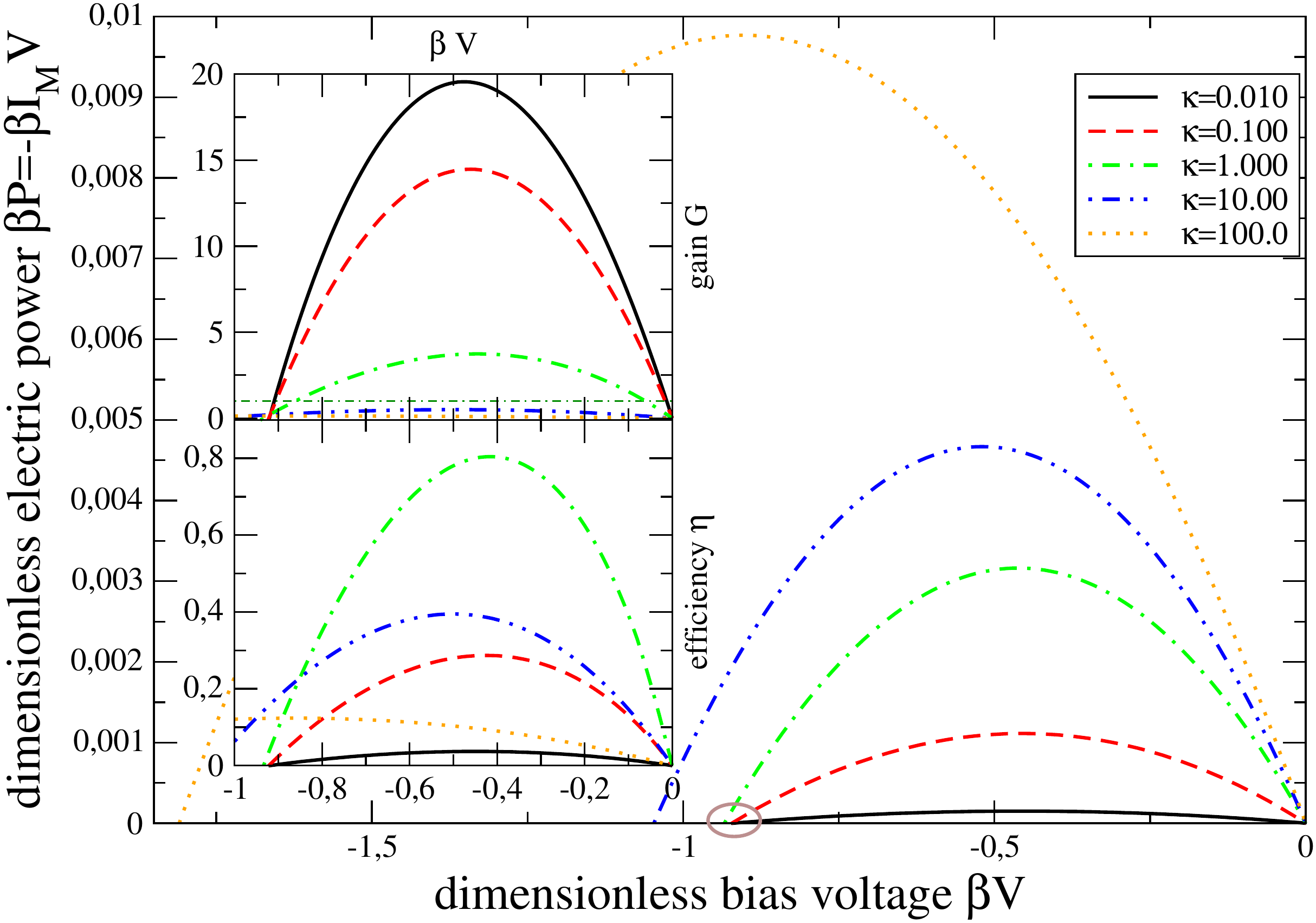}
\caption{\label{FIG:power_vs_coupling}
Plot of the electric power vs. bias voltage.
With increasing coupling strength the generated power as well as the window of positive power increases, but so does the 
energy consumption needed to run the feedback loop.
Top Inset: The gain (similar color coding) defined in Eq.~(\ref{EQ:gain}) demonstrates that in the strong coupling regime, 
more energy is needed to run the engine than is gained from the produced power, 
manifest in a gain factor below one (thin dash-dotted line).
Bottom Inset: The efficiency defined in Eq.~(\ref{EQ:efficiency}) is not a monotonous function of the coupling strength, and at intermediate coupling strengths 
both information and feedback energy are best converted into electric power.
Parameters for $\kappa=0.01$ are identical with Fig.~\ref{FIG:feedback_effect}, other curves only have correspondingly scaled coupling constants.
}
\end{figure*}

A naive extrapolation of the SQD treatment towards the strong coupling limit predicts that all currents and derived quantities such as generated
power should scale linearly in the coupling strength, apparently predicting no limit in power production.
However, from the exact solution of the SQD in absence of feedback control we know that by increasing the coupling strength to the reservoirs, 
the current through the system can be increased only up to a finite limit~\cite{topp2015a}, 
see also the benchmark in the companion paper~\cite{strasberg2018a}.
Therefore, it is an intriguing question how the generated electric power scales in the strong-coupling regime.
In this section, we therefore investigate how the currents change when the coupling strength is scaled by a factor $\Gamma_\alpha^{{\rm E }/{\rm F}} \to \kappa \Gamma_\alpha^{{\rm E }/{\rm F}}$.
We note that this convention leads to larger differences between the coupling constants and thereby also to a larger switching work as $\kappa$ is increased.
In Fig.~\ref{FIG:power_vs_coupling} we show the power vs. bias voltage for different coupling strengths, where we adopt the convention that
the previous parameters of the weak-coupling limit (Fig.~\ref{FIG:feedback_effect}) are reproduced when $\kappa=0.01$.
We observe that by increasing the coupling strength, the electric output power is indeed increased as well (main plot).
For weak to moderate coupling strength (solid black, dashed red, and dash-dotted green curves), we see that the device is still information-dominated, 
as the gain factor~(\ref{EQ:gain}) can become larger than one, indicating that then mainly information is converted to electric output power.
However, as the gain factor $G$ continuously decreases with increasing coupling strength, beyond a critical coupling, the device is no longer 
information-dominated, and the gain $G$ is smaller than one (dot-dot-dashed blue and dotted brown curves in the top inset of Fig.~\ref{FIG:power_vs_coupling}).
Considering both information and feedback energy as consumed resources, the efficiency~(\ref{EQ:efficiency}) becomes the relevant figure of merit.
In contrast to the gain, the observed maximum efficiency does not evolve monotonously with the coupling strength (bottom inset).
It first grows with increasing coupling strength, and we observe an acceptable 
maximum efficiency of nearly $80\%$ at a moderate coupling strength where 
the device is still information-dominated (green dash-dotted curve).
However, for stronger and ultrastrong couplings, the efficiency decreases again (dot-dot-dashed blue and dotted brown curves), 
such that in this energy-dominated regime, the device is useless from a practical viewpoint:
Running the feedback loop requires more energy than is generated from information.

Additionally, we observe that in the information-dominated (Maxwell-demon) regime, where the gain is larger than one, also the non-trivial point $V^*$ (brown circle), 
where the generated power vanishes, is hardly dependent on the coupling strength.
Experimentally, this may be an important hallmark for the identification of this regime.
For the naive SQD treatment, the position of this point may be calculated analytically
\bea
\beta V^* = -\ln \frac{\Gamma_{\rm L}^{\rm E} \Gamma_{\rm R}^{\rm F}}{\Gamma_{\rm L}^{\rm F} \Gamma_{\rm R}^{\rm E}}\,,
\eea
which is however significantly larger than the observed value for the TQD treatment even in the weak-coupling regime, compare also the main plot of Fig.~\ref{FIG:feedback_effect}.
We attribute this to the inherent weakness of the measurement, which strongly delimits the capabilities of the demon.

%%%%%%%%%%%%%%%%%%%%%%%%%%%%%%%%%%%%%%%%%%%%%%%%%%%%%%%%%%%%%%%%%%%%%%%%%%%%%%%%%%%%%%%%%%%%%%%%%%%
%%%%%%%%%%%%%%%%%%%%%%%%%%%%%%%%%%%%%%%%%%%%%%%%%%%%%%%%%%%%%%%%%%%%%%%%%%%%%%%%%%%%%%%%%%%%%%%%%%%

\subsection{Coarse-graining effects}\label{SEC:coarsegraining}

We see that in the TQD treatment, the electric power produced by the device is significantly smaller than in the SQD treatment, compare
Fig.~\ref{FIG:feedback_effect}.
Consistently, the efficiency in the weak-coupling regime (where the device operates information-dominated) is also significantly below
the SQD efficiency, compare Fig.~\ref{FIG:power_vs_coupling} with the discussion in App.~\ref{SEC:sqd_efficiency}.
This leads us to the conclusion that the presented TQD treatment does not efficiently use the information to close the feedback loop.
Knowing that coarse-graining strongly influences the entropy balance~\cite{esposito2012b}, 
one might question whether this results from the employed coarse-graining of measurement outcomes into ${\cal M}_{\rm E}$ and ${\cal M}_{\rm F}$.
Instead of just using the two coarse-grained rates $\Gamma_\alpha^{{\rm E }/{\rm F}}$, we can consider $n$-conditioned rates, where a suitable choice could be
\bea
\Gamma_\alpha^{(n)} = \frac{x^n e^{-x} \Gamma_\alpha^{\rm E} + y^n e^{-y} \Gamma_\alpha^{\rm F}}{x^n e^{-x} + y^n e^{-y}}\,.
\eea
This will preserve positive tunneling rates throughout with similar tunneling rates as in the coarse-grained picture.
Furthermore, if $y < n < x$, the conditional rates will be between the coarse-grained ones, 
and in particular $\Gamma_\alpha^{(n_{\rm thr})} = \frac{1}{2}(\Gamma_\alpha^{\rm E} + \Gamma_\alpha^{\rm F})$,
such that the feedback is adapted to the weakness of the measurement.
The information gained by the measurement is thus no longer discarded.
We have observed however, that the generated electric power is not significantly enhanced by this procedure, see the dash-dotted magenta curve in Fig.~\ref{FIG:feedback_effect}.

%%%%%%%%%%%%%%%%%%%%%%%%%%%%%%%%%%%%%%%%%%%%%%%%%%%%%%%%%%%%%%%%%%%%%%%%%%%%%%%%%%%%%%%%%%%%%%%%%%%
%%%%%%%%%%%%%%%%%%%%%%%%%%%%%%%%%%%%%%%%%%%%%%%%%%%%%%%%%%%%%%%%%%%%%%%%%%%%%%%%%%%%%%%%%%%%%%%%%%%
%%%%%%%%%%%%%%%%%%%%%%%%%%%%%%%%%%%%%%%%%%%%%%%%%%%%%%%%%%%%%%%%%%%%%%%%%%%%%%%%%%%%%%%%%%%%%%%%%%%
%%%%%%%%%%%%%%%%%%%%%%%%%%%%%%%%%%%%%%%%%%%%%%%%%%%%%%%%%%%%%%%%%%%%%%%%%%%%%%%%%%%%%%%%%%%%%%%%%%%

\section{Summary}

We have considered the performance of an externally controlled feedback loop implementing an electronic Maxwell demon.
To explore the strong-coupling limit, we employed a fermionic reaction-coordinate mapping to an effective
triple quantum dot system, serially coupled to two leads.
Combining this with continuous projective measurements of the central dot occupation, the destruction of coherent superpositions
within the triple dot system led to a complete suppression of the current due to the quantum Zeno effect.
Since the mapping holds also for weak couplings, this raises the question why the Zeno suppression was not observed in the
original approach based on a single dot rate equation~\cite{schaller2011b}.
An independent investigation shows that this is due to the inherent Markovian assumption in the single dot rate equation:
If a non-Markovian approach is applied to the single-dot feedback problem, the Zeno suppression is found~\cite{engelhardt2017c}.
By performing a mapping to a triple quantum dot, we obtain a Markovian embedding, which captures the non-Markovian Zeno suppression
in the single quantum dot.

A Zeno suppression is not directly observed in the numerous counting statistics experiments, with detectors always switched
on and thereby continuously measuring.
We therefore are led to believe that realistic charge measurements are far from projective, and reflected this
in our work by generalizing our model to weak measurements.
Inspired by charge detectors used in experiments, we implemented this by using a microscopic detector model for a point contact.
Effectively, this led already in the weak-coupling limit to an overall reduced performance of the demon in comparison 
with the original single-dot model -- both in terms of information-to-power conversion
efficiency and electric power output.
In the intermediate coupling strength regime, the energy to run the feedback loop becomes important, and the driving force is no longer information
but the device rather acts like a pump. 
Finally, in the strong-coupling regime, the energetic contribution (opening and closing of the shutter) becomes dominant, showing
that the device can no longer be interpreted as a demon.

Beyond obvious applications to more advanced models (e.g. Coulomb interactions, spin valves), it would in the future 
also be interesting to discuss the implications of finite-time control cycles, which enables to lift the strong-coupling assumption during detection.
Then, measurements may be constructed that leave the system energy invariant, such that the only energy required for the feedback loop is the switching work.

%%%%%%%%%%%%%%%%%%%%%%%%%%%%%%%%%%%%%%%%%%%%%%%%%%%%%%%%%%%%%%%%%%%%%%%%%%%%%%%%%%%%%%%%%%%%%%%%%%%
%%%%%%%%%%%%%%%%%%%%%%%%%%%%%%%%%%%%%%%%%%%%%%%%%%%%%%%%%%%%%%%%%%%%%%%%%%%%%%%%%%%%%%%%%%%%%%%%%%%
%%%%%%%%%%%%%%%%%%%%%%%%%%%%%%%%%%%%%%%%%%%%%%%%%%%%%%%%%%%%%%%%%%%%%%%%%%%%%%%%%%%%%%%%%%%%%%%%%%%
%%%%%%%%%%%%%%%%%%%%%%%%%%%%%%%%%%%%%%%%%%%%%%%%%%%%%%%%%%%%%%%%%%%%%%%%%%%%%%%%%%%%%%%%%%%%%%%%%%%

\section{Acknowledgments}

Financial support by the DFG (grants SCHA 1646/3-1, SFB 910, GRK 1558), the European Research Council (Project No. 681456), 
the WE-Heraeus foundation (WEH 640) as well as stimulating discussions with 
A. Carmele, S. Gurvitz, A. Nazir, and D. Segal are gratefully acknowledged.
The authors are indebted to Tobias Brandes for motivating the research.

%%%%%%%%%%%%%%%%%%%%%%%%%%%%%%%%%%%%%%%%%%%%%%%%%%%%%%%%%%%%%%%%%%%%%%%%%%%%%%%%%%%%%%%%%%%%%%%%%%%
%%%%%%%%%%%%%%%%%%%%%%%%%%%%%%%%%%%%%%%%%%%%%%%%%%%%%%%%%%%%%%%%%%%%%%%%%%%%%%%%%%%%%%%%%%%%%%%%%%%
%%%%%%%%%%%%%%%%%%%%%%%%%%%%%%%%%%%%%%%%%%%%%%%%%%%%%%%%%%%%%%%%%%%%%%%%%%%%%%%%%%%%%%%%%%%%%%%%%%%
%%%%%%%%%%%%%%%%%%%%%%%%%%%%%%%%%%%%%%%%%%%%%%%%%%%%%%%%%%%%%%%%%%%%%%%%%%%%%%%%%%%%%%%%%%%%%%%%%%%

\bibliographystyle{apsrev4-1}
\bibliography{/home/schaller/literatur/postdoc/postdoc}

%%%%%%%%%%%%%%%%%%%%%%%%%%%%%%%%%%%%%%%%%%%%%%%%%%%%%%%%%%%%%%%%%%%%%%%%%%%%%%%%%%%%%%%%%%%%%%%%%%%
%%%%%%%%%%%%%%%%%%%%%%%%%%%%%%%%%%%%%%%%%%%%%%%%%%%%%%%%%%%%%%%%%%%%%%%%%%%%%%%%%%%%%%%%%%%%%%%%%%%
%%%%%%%%%%%%%%%%%%%%%%%%%%%%%%%%%%%%%%%%%%%%%%%%%%%%%%%%%%%%%%%%%%%%%%%%%%%%%%%%%%%%%%%%%%%%%%%%%%%
%%%%%%%%%%%%%%%%%%%%%%%%%%%%%%%%%%%%%%%%%%%%%%%%%%%%%%%%%%%%%%%%%%%%%%%%%%%%%%%%%%%%%%%%%%%%%%%%%%%

\appendix

%%%%%%%%%%%%%%%%%%%%%%%%%%%%%%%%%%%%%%%%%%%%%%%%%%%%%%%%%%%%%%%%%%%%%%%%%%%%%%%%%%%%%%%%%%%%%%%%%%%
%%%%%%%%%%%%%%%%%%%%%%%%%%%%%%%%%%%%%%%%%%%%%%%%%%%%%%%%%%%%%%%%%%%%%%%%%%%%%%%%%%%%%%%%%%%%%%%%%%%

\section{Reaction Coordinate mapping from SQD to TQD}\label{APP:reaction_coordinate}

Given a spectral density $\Gamma^{(0)}(\omega)$ for the initial reservoir, 
the new coupling strength $\lambda$ between system and reaction coordinate and 
the energy $\Omega$ of the reaction coordinate are calculated according to (compare
also Ref.~\cite{strasberg2018a})
\bea
\abs{\lambda}^2 &=& \frac{1}{2\pi} \int \Gamma^{(0)}(\omega) d\omega\,,\nn
\Omega &=& \frac{1}{2\pi\abs{\lambda}^2} \int \omega\Gamma^{(0)}(\omega) d\omega\,.
\eea
We note that the transformation does not determine the phase of $\lambda$.
This phase can be transformed away, such that we will not consider it here.
Finally, the new spectral density can be obtained from the old one via
\bea
\Gamma^{(1)}(\omega) = \frac{4 \abs{\lambda}^2 \Gamma^{(0)}(\omega)}{\left[\frac{1}{\pi} {\cal P}\int \frac{\Gamma^{(0)}(\omega')}{\omega'-\omega} d\omega'\right]^2 + \left[\Gamma^{(0)}(\omega)\right]^2}\,,
\eea
where ${\cal P}$ denotes the principal value.
These equations can be derived from the Heisenberg picture dynamics of the annihilation operators of the central dot~\cite{strasberg2018a}.
When we introduce a reaction coordinate for both left and right leads ($\lambda\to\lambda_\alpha$ and $\Omega\to\Omega_\alpha$), we obtain
for the Lorentzian density (\ref{EQ:spectral_density_old}) the results
$\abs{\lambda_\alpha}^2 = \frac{\Gamma_\alpha \delta_\alpha}{2}$ and
$\Omega_\alpha = \epsilon_\alpha$.
Altogether, this implies the mapping~(\ref{EQ:ham_mapped}) with Eq.~(\ref{EQ:spectral_density_new}) exemplified in the main article.
We also mention that to apply such mappings recursively ad infinitum, it would be 
necessary that the spectral density has a rigid cutoff.
In contrast, for the considered example a repeated application of the mapping would lead to divergent integrals and will not be performed.

%%%%%%%%%%%%%%%%%%%%%%%%%%%%%%%%%%%%%%%%%%%%%%%%%%%%%%%%%%%%%%%%%%%%%%%%%%%%%%%%%%%%%%%%%%%%%%%%%%%
%%%%%%%%%%%%%%%%%%%%%%%%%%%%%%%%%%%%%%%%%%%%%%%%%%%%%%%%%%%%%%%%%%%%%%%%%%%%%%%%%%%%%%%%%%%%%%%%%%%

\section{Derivation of the measurement superoperator}\label{APP:measurement}

%%%%%%%%%%%%%%%%%%%%%%%%%%%%%%%%%%%%%%%%%%%%%%%%%%%%%%%%%%%%%%%%%%%%%%%%%%%%%%%%%%%%%%%%%%%%%%%%%%%
\subsection{Measurement dissipator without counting fields}

We start from an interaction Hamiltonian between central dot and the QPC of the form
\bea
H_I = A \sum_{kk'} \left[ t_{kk'} \gamma_{k{\rm L}} \gamma_{k'{\rm R}}^\dagger + {\rm h.c.}\right]\,,\;\;
A = \f{1}-\sigma d^\dagger d\,,\qquad
\eea
where the system coupling operator $A$ suppresses the tunneling amplitudes $t_{kk'}$ between left and right modes of the QPC 
$H_{QPC} = \sum_k \epsilon_{k{\rm L}} \gamma_{k{\rm L}}^\dagger \gamma_{k{\rm L}} + \sum_k \epsilon_{k{\rm R}} \gamma_{k{\rm R}}^\dagger \gamma_{k{\rm R}}$ by $1-\sigma$ when the central dot
of the TQD system is occupied.
Thus, when $\sigma=0$, the QPC is insensitive to the charge of the central dot, and for $\sigma=1$, QPC transport is completely blocked when the central dot is filled.
We now assume in addition that during the measurement, the interaction dominates the internal dynamics of the TQD system.
Then, the singular coupling limit is applicable (compare e.g. Sec. 3.3.3 in Ref.~\cite{breuer2002}), which automatically leads to a dissipator of Lindblad 
form.
This will locally act only on the central dot, just as if it was only an SQD coupled to the QPC.
In the limit where the bias voltage of the QPC is large $\beta V_{\rm QPC}\gg 1$ we can consider the wideband limit (compare Sec.~5.4 of Ref.~\cite{schaller2014})
where $T(\omega,\omega')\equiv 2 \pi \sum_{kk'} \abs{t_{kk'}}^2 \delta(\omega-\epsilon_{k{\rm L}})\delta(\omega'-\epsilon_{k'{\rm R}}) \to T_0$, which condenses into
the QPC tunneling rate $\gamma=T_0 V_{\rm QPC}$ and the simple dissipator
\bea
{\cal L}_{\rm dt}\rho &=& \gamma \left[A \rho A^\dagger - \frac{1}{2} \left\{A^\dagger A, \rho\right\}\right]\nn
&&= \gamma \sigma^2 \left[d^\dagger d \rho d^\dagger d - \frac{1}{2} \left\{d^\dagger d, \rho\right\}\right]\,.
\eea
In the last line, we have inserted the definition of $A$ and used the fermionic anticommutation relations.
This particularly simple form allows one to easily compute the exponential of ${\cal L}_{\rm dt}$.
Defining the superoperators ${\cal J} \rho = d^\dagger d \rho d^\dagger d$, ${\cal J}_{\rm L} \rho = \frac{1}{2} d^\dagger d \rho$, and ${\cal J}_{\rm R} = \frac{1}{2} \rho d^\dagger d$, we can --
since they all mutually commute -- compute their action separately
$e^{-\gamma\sigma^2 {\cal J}_{\rm L} \Delta\tau}\rho = \left[d d^\dagger + e^{-\sigma^2 \gamma\Delta\tau/2} d^\dagger d\right] \rho$,
$e^{-\gamma\sigma^2 {\cal J}_{\rm R} \Delta\tau}\rho = \rho \left[d d^\dagger + e^{-\sigma^2 \gamma\Delta\tau/2} d^\dagger d\right]$,
and $e^{+\gamma \sigma^2 {\cal J} \Delta\tau}\rho = \rho + \left(e^{+\sigma^2 \gamma\Delta\tau}-1\right) d^\dagger d \rho d^\dagger d$,
which upon sequential application yields
for the total effect of the measurement on average
\bea
e^{{\cal L}_{\rm dt}(0)\Delta\tau} \rho &=& d d^\dagger \rho d^\dagger d + d^\dagger d \rho d^\dagger d\\
&&+ e^{-\sigma^2 \gamma\Delta\tau/2} \left(d^\dagger d \rho d d^\dagger + d d^\dagger \rho d^\dagger d\right)\,.\nonumber
\eea
From this, we can see that on average the effect of the measurement is just the destruction of coherences between empty and filled central dot states.
By employing the fermionic anticommutation relations we therefore get the expression
\bea
\frac{e^{{\cal L}_{\rm dt}(0)\Delta\tau}-\f{1}}{\Delta t} 
= \frac{e^{-\sigma^2\frac{\gamma\Delta\tau}{2}}-1}{\Delta t} \left(d^\dagger d \rho d d^\dagger + {\rm h.c.}\right)\,,
\eea
which implies with ${\cal M}_E + {\cal M}_F = e^{{\cal L}_{\rm dt}(0)\Delta\tau}$ -- see the discussion in the subsection below -- Eq.~(\ref{EQ:dissipator_measurement}) in the article.

%%%%%%%%%%%%%%%%%%%%%%%%%%%%%%%%%%%%%%%%%%%%%%%%%%%%%%%%%%%%%%%%%%%%%%%%%%%%%%%%%%%%%%%%%%%%%%%%%%%
\subsection{Measurement dissipator in presence of QPC electron counting}

To furthermore infer the counting statistics of the QPC, it is a well-established practice to introduce counting fields~\cite{esposito2009a}, 
which yields a generalized dissipator of the form
\bea\label{EQ:dissipator_qpc}
{\cal L}_{\rm dt}(\chi) \rho = \gamma \left[e^{+\ii\chi} A \rho A^\dagger - \frac{1}{2} \left\{A^\dagger A, \rho\right\}\right]\,,
\eea
where $\chi$ denotes the counting field for the charges transferred through the QPC circuit.
By computing derivatives of the moment-generating function $M(\chi) = \trace{e^{{\cal L}_{\rm dt}(\chi) \Delta\tau} \rho(t)}$ with 
respect to the counting field $\chi$, we can determine all moments of the charge distributions of tunneled QPC charges during the interval $[t,t+\Delta\tau]$,
where $\Delta \tau$ denotes the duration of the measurement.
By construction, the inverse Fourier transform  of the generating function yields the full distribution
\bea
P_n(\Delta \tau) = \trace{\frac{1}{2\pi} \int_{-\pi}^{+\pi} e^{{\cal L}_{\rm dt}(\chi) \Delta\tau - \ii n \chi} \rho(t)}\,,
\eea
and the corresponding conditional (not normalized) density matrix is given by~\cite{schaller2017a}
\bea
\rho^{(n)}(t+\Delta \tau) &=& \frac{1}{2\pi} \int_{-\pi}^{+\pi} e^{{\cal L}_{\rm dt}(\chi) \Delta\tau - \ii n \chi} d\chi \rho(t)\nn
&\equiv& {\cal M}_n \rho(t)\,,
\eea
which defines the measurement superoperators ${\cal M}_n$.
In system (TQD)-detector (QPC transfer particle number) Hilbert space the most general density matrix can be written as
$\rho_{\rm SD}(t) = \sum_{nm} \rho^{(nm)}(t) \otimes \ket{n}\bra{m}$, such that by performing a projective measurement of the number of 
particles transferred through the QPC and trace out the detector afterwards, this leads to the identification
$\rho^{(n)} = \rho^{(nn)}$.
We note that from the completeness relation of the Fourier transform we can also infer that 
$\rho(t+\Delta \tau) = \sum_n \rho^{(n)}(t+\Delta\tau) = e^{{\cal L}_{\rm dt}(0) \Delta\tau} \rho(t)$, such that $\sum_n {\cal M}_n = e^{{\cal L}_{\rm dt}(0)\Delta\tau}$.
In a similar fashion as before, we can also partition the generalized dissipator into mutually commuting superoperators
${\cal L}_{\rm dt}(\chi) = \gamma \left[e^{+\ii\chi} {\cal J} + {\cal J}_{\rm L} + {\cal J}_{\rm R}\right]$, for which we can
separately compute the exponential.
Eventually, this yields
\bea
{\cal M}_n &=& \frac{(\gamma\Delta\tau)^n}{n!}{\cal J}^n e^{-\gamma\Delta\tau ({\cal J}_{\rm L} + {\cal J}_{\rm R})}\,.
\eea
By using that 
$A^n = d d^\dagger + (1-\sigma)^n d^\dagger d$
we can explicitly determine the individual superoperators
\bea
{\cal J}^n \rho &\hat{=}& \left[d d^\dagger + (1-\sigma)^n d^\dagger d\right] \rho\times\\
&&\times\left[d d^\dagger + (1-\sigma)^n d^\dagger d\right]\,,\nn
e^{-\gamma\Delta\tau ({\cal J}_{\rm L} + {\cal J}_{\rm R})} \rho &\hat{=}& \left[e^{-\gamma\Delta\tau/2} d d^\dagger + e^{-\gamma\Delta\tau(1-\sigma)^2/2} d^\dagger d\right] \rho\times\nn
&&\times\left[e^{-\gamma\Delta\tau/2} d d^\dagger + e^{-\gamma\Delta\tau(1-\sigma)^2/2} d^\dagger d\right]\,,\nonumber
\eea
which upon nested application eventually leads to Eq.~(\ref{EQ:povm_measurement}) in the main article.

%%%%%%%%%%%%%%%%%%%%%%%%%%%%%%%%%%%%%%%%%%%%%%%%%%%%%%%%%%%%%%%%%%%%%%%%%%%%%%%%%%%%%%%%%%%%%%%%%%%
\subsection{Example: SQD monitored by QPC}\label{APP:example_set_qpc}

Taking ${\cal L}_{\rm dt}(\chi)$ from Eq.~(\ref{EQ:dissipator_qpc}) with $A=\f{1}-\sigma d^\dagger d$, we can evaluate the action of
the QPC in the dot eigenbasis $\ket{0}$, $\ket{1}$.
Due to the special form of the dissipator, it does not couple between the populations and the coherences of the dot density matrix, and 
we get
$\left({\cal L}_{\rm dt}(\chi) \rho\right)_{00} = \gamma (e^{+\ii\chi}-1) \rho_{00}$ and 
$\left({\cal L}_{\rm dt}(\chi) \rho\right)_{11} = \gamma (1-\sigma)^2 (e^{+\ii\chi}-1) \rho_{11}$.
Since for a single quantum dot system, coherences do not play a role, we therefore obtain that the generalized rate matrix
(acting on the probability vector $\f{P} = (p_{\rm E}, p_{\rm F})^{\rm T}$ for finding the dot empty or filled, respectively) is diagonal in the SQD eigenbasis
\bea
{\cal W}_{\rm dt}(\chi) = \gamma (e^{+\ii\chi}-1) \left(\begin{array}{cc}
1 & 0\\
0 & (1-\sigma)^2
\end{array}\right)\,.
\eea
Here, we see that for the chosen large QPC-bias limit, we only have unidirectional transport through the QPC, such that only
terms with $e^{+\ii \chi}$ occur, and $\gamma>0$ describes the QPC transmission (compare Sec.~5.4.1 of Ref.~\cite{schaller2014}).
The total rate matrix is then constructed additively -- compare Eq.~(\ref{EQ:rateeq_set}) -- which leads to
\bea
{\cal W}(\chi) = {\cal W}_{\rm dt}(\chi) + {\cal W}_L + {\cal W}_R\,.
\eea
Here, the rate matrices ${\cal W}_{L/R}$ describe electronic jumps onto or off the dot from the left and right leads, respectively, 
and ${\cal W}_{\rm dt}$ is a generalized rate matrix for the QPC, with the counting field $\chi$ describing the number of transferred
QPC charges.
In absence of counting ($\chi=0$), the effect of the QPC on the single dot vanishes, for larger system such as e.g. a double dot however, 
the QPC would still have an effect~\cite{bulnes_cuetara2013a}.

The above rate matrix is an extremely simple example of a bistable stochastic process: For vanishing SQD tunneling rates ${\cal W}_{L/R}\to 0$, 
the dot occupation cannot change, and the statistics $P_n(\Delta t)$ will be fully Poissonian, depending on only the initial SQD occupation:
When it is empty, the cumulants will be given by $\gamma \Delta t$, and when it is filled, they are given by $\gamma(1-\sigma)^2\Delta t$.
The interesting case arises when the SQD rate matrices ${\cal W}_{L/R}$ are small in comparison to ${\cal W}_{\rm dt}$: Then, slow switching 
events occur between the two Poissonian distributions~\cite{jordan2004a}, and the time-dependent detector signal can be used to infer the 
occupation of the dot~\cite{flindt2009a}.

From the theory of Full Counting Statistics, we can infer the probability $P_n(\Delta t)$ of observing $n$ QPC charge transfer events
during in the interval $[t,t+\Delta t]$ via
\bea
P_n(\Delta t) = \frac{1}{2\pi} \int_{-\pi}^{+\pi} \trace{e^{{\cal W}(\chi) \Delta t} \f{P}(t)} e^{-\ii n \chi} d\chi\,,
\eea
where the trace corresponds in this case to the multiplication from the left with the row vector $(1,1)$.
Furthermore, a measurement of $n$ QPC charge transfers after $\Delta t$ would project the probability vector to
\bea
\f{P}^{(n)}(t+\Delta t) = \frac{1}{2\pi} \int_{-\pi}^{+\pi} e^{{\cal W}(\chi) \Delta t} \f{P}(t) e^{-\ii n \chi} d\chi\,,
\eea
which still needs to be normalized by $P_n(\Delta t)$.
In the limit where $\gamma \gg \Gamma_\alpha$, a perturbative treatment for these expressions can be used.
To generate a trajectory such as in Fig.~\ref{FIG:cvecqpc_measurement}, we start with an empty dot $\f{P}(0) = (1,0)$,
then compute the probabilities $P_n(\Delta t)$, choose accordingly
a particular outcome $n$ -- which defines a current $I=n/\Delta t$ -- 
and perform the projection, which leads to $\f{P}(\Delta t)$. 
This is then taken as the initial state for the next iteration and so on.
Due to the projection, it is more likely to measure large currents after large currents and low currents after low currents, which
leads to the switching behaviour shown in Fig.~\ref{FIG:cvecqpc_measurement}.

%%%%%%%%%%%%%%%%%%%%%%%%%%%%%%%%%%%%%%%%%%%%%%%%%%%%%%%%%%%%%%%%%%%%%%%%%%%%%%%%%%%%%%%%%%%%%%%%%%%
%%%%%%%%%%%%%%%%%%%%%%%%%%%%%%%%%%%%%%%%%%%%%%%%%%%%%%%%%%%%%%%%%%%%%%%%%%%%%%%%%%%%%%%%%%%%%%%%%%%

\section{Triple dot properties}\label{APP:tripledot}

\subsection{Spectrum}\label{APP:tripledot_spectrum}

In the simple case when $\epsilon_{\rm L}=\epsilon_{\rm R}=\epsilon$, the spectrum and the eigenvectors of the TQD part
%\bea
%H_{\rm S} = \epsilon_{\rm L} d_{\rm L}^\dagger d_{\rm L} + \epsilon d^\dagger d + \epsilon_{\rm R} d_{\rm R}^\dagger d_{\rm R} + \sqrt{\frac{\Gamma_{\rm L} \delta_{\rm L}}{2}} \left(d_{\rm L} d^\dagger + d d_{\rm L}^\dagger\right)
%+ \sqrt{\frac{\Gamma_{\rm R} \delta_{\rm R}}{2}} \left(d_{\rm R} d^\dagger + d d_{\rm R}^\dagger\right)
%\eea
of the Hamiltonian~(\ref{EQ:ham_mapped}) can be computed analytically.
The eigenstates can be grouped into states with the same total particle number, with energies
\bea
E_0 &=& 0\,,\nn
E_1^0 &=& \epsilon\,,\qquad
E_1^\pm = \epsilon \pm \sqrt{\frac{\Gamma_{\rm L} \delta_{\rm L}}{2} +\frac{\Gamma_{\rm R} \delta_{\rm R}}{2}}\,,\nn
E_2^0 &=& 2\epsilon\,,\qquad
E_2^\pm = 2\epsilon\pm\sqrt{\frac{\Gamma_{\rm L} \delta_{\rm L}}{2} +\frac{\Gamma_{\rm R} \delta_{\rm R}}{2}}\,,\nn
E_3 &=& 3\epsilon\,.
\eea

We note that the TQD energies become near degenerate for small $\Gamma_\alpha \delta_\alpha$.
Applying the Master equation formalism to it should be well justified now when $\delta_\alpha \beta \ll 1$.
Applying the secular approximation on top however should only be admissible when $\sqrt{\Gamma_\alpha \delta_\alpha} \gg \delta_\alpha$.

Furthermore, it should be noted that out of the 64 matrix elements of the TQD density matrix, not all are allowed within our treatment, as a master equation treatment of the TQD
will only admit to create superpositions of states of similar charge on the TQD.
That is, we can have coherences between the singly and the doubly charged states separately, leading to $20=1+9+9+1$ non-vanishing density matrix elements in total.
Taking only these physically allowed matrix elements into account and then performing the partial trace over the left and right dots, we obtain that the
reduced density matrix of the central dot must always be diagonal.

%Further, when the tunneling rates $\Gamma_\alpha$ become time-dependent, two different modes of operation become visible:
%
%For example, when $\Gamma_\alpha^{{\rm E }/{\rm F}}= \alpha^{{\rm E}/{\rm F}} \Gamma_\alpha$, the effect of the feedback will not change the eigenvectors, and only the spectrum is changed.
%
%Second, when $\Gamma_{\rm L}^{\rm E} \delta_{\rm L} + \Gamma_{\rm R}^{\rm E} \delta_{\rm R} = \Gamma_{\rm L}^{\rm F} \delta_{\rm L} + \Gamma_{\rm R}^{\rm F} \delta_{\rm R}$, the spectrum remains unchanged, but only the eigenvectors change.

%%%%%%%%%%%%%%%%%%%%%%%%%%%%%%%%%%%%%%%%%%%%%%%%%%%%%%%%%%%%%%%%%%%%%%%%%%%%%%%%%%%%%%%%%%%%%%%%%%%

\subsection{Correlation functions}

Identifying the coupling operators between the TQD and the residual reservoirs as
\bea
A_1 &=& d_{\rm L}^\dagger\,,\qquad B_1 = \sum_k T_{k{\rm L}}^* C_{k{\rm L}}\,,\nn
A_2 &=& d_{\rm L}\,,\qquad B_2 = \sum_k T_{k{\rm L}} C_{k{\rm L}}^\dagger\,,\nn
A_3 &=& d_{\rm R}^\dagger\,,\qquad B_3 = \sum_k T_{k{\rm R}}^* C_{k{\rm R}}\,,\nn
A_4 &=& d_{\rm R}\,,\qquad B_4 = \sum_k T_{k{\rm R}} C_{k{\rm R}}^\dagger\,,
\eea
we can represent the non-vanishing correlation functions as (compare e.g. Chapter~5 in Ref.~\cite{schaller2014})
\bea
C_{12}(\tau) &=& \frac{1}{2\pi} \int\limits_{-\infty}^{+\infty} \Gamma_{\rm L}^{(1)}(\omega) [1-f_{\rm L}(\omega)] e^{-\ii\omega \tau} d\omega\,,\nn
C_{21}(\tau) &=& \frac{1}{2\pi} \int\limits_{-\infty}^{+\infty} \Gamma_{\rm L}^{(1)}(\omega) f_{\rm L}(\omega) e^{+\ii\omega \tau} d\omega\,,
%C_{34}(\tau) &=& \frac{1}{2\pi} \int\limits_{-\infty}^{+\infty} \Gamma_{\rm R}^{(1)}(\omega) [1-f_{\rm R}(\omega)] e^{-\ii\omega \tau} d\omega\,,\qquad
%C_{43}(\tau) = \frac{1}{2\pi} \int\limits_{-\infty}^{+\infty} \Gamma_{\rm R}^{(1)}(\omega) f_{\rm R}(\omega) e^{+\ii\omega \tau} d\omega\,.
\eea
and similar for $C_{34}(\tau)$ and $C_{43}(\tau)$ by replacing L$\to$R.
From this, we can read off the Fourier transform of e.g. the left-associated correlation functions
\bea
\gamma_{12}(\omega) &=& \Gamma_{\rm L}^{(1)}(+\omega)[1-f_{\rm L}(+\omega)]\,,\nn
\gamma_{21}(\omega) &=& \Gamma_{\rm L}^{(1)}(-\omega)f_{\rm L}(-\omega)\,.
\eea

%%%%%%%%%%%%%%%%%%%%%%%%%%%%%%%%%%%%%%%%%%%%%%%%%%%%%%%%%%%%%%%%%%%%%%%%%%%%%%%%%%%%%%%%%%%%%%%%%%%
\subsection{Born-Markov master equation in absence of feedback}\label{APP:bornmarkovme}

We can decide not to perform the secular approximation, but only the Born and Markov approximations. 
This will in general not lead to a Lindblad type master equation
\bea
\dot{\rho} = {\cal L}\rho = -\ii \left[H_{\rm S}, \rho\right] + \sum_\alpha {\cal L}^{(\alpha)} \rho\,,
\eea
but we can nevertheless expect that for weak residual couplings $\delta_\alpha$ it will approximately preserve the basic thermodynamic properties of the system.

In our case, it assumes the form
\bea\label{EQ:master_markov}
\dot{\rho} &=& -\ii [H_{\rm S}, \rho]\nn
&&-\left\{ [c_{\rm L}, M_{{\rm L},21} \rho] +[c_{\rm L}^\dagger, M_{{\rm L},12} \rho] + {\rm h.c.}\right\}\nn
&&-\left\{ [c_{\rm R}, M_{{\rm R},21} \rho] +[c_{\rm R}^\dagger, M_{{\rm R},12} \rho] + {\rm h.c.}\right\}\,,
\eea
where (similar for the right lead L$\to$R and $12\to 34$)
\bea
M_{{\rm L},21} &=& \sum_{ab} \Gamma_{21}(E_b-E_a) \bra{a} c_{\rm L}^\dagger \ket{b} \ket{a} \bra{b}\,,\nn
M_{{\rm L},12} &=& \sum_{ab} \Gamma_{12}(E_b-E_a) \bra{a} c_{\rm L} \ket{b} \ket{a} \bra{b}\,.
\eea
Here, we have used the TQD energy eigenbasis $H_S \ket{a} = E_a \ket{a}$ (compare App.~\ref{APP:tripledot_spectrum}) and 
the half-sided Fourier transform of the correlation function
\bea
\Gamma_{\alpha\beta}(\omega) = \int_0^\infty C_{\alpha\beta}(\tau) e^{+\ii\omega\tau} d\tau\,.
\eea
These can be rewritten using the convolution theorem
\bea
\Gamma_{\alpha\beta}(\omega) &=& \int\limits_{-\infty}^{+\infty} C_{\alpha\beta}(\tau) \Theta(\tau) e^{+\ii\omega\tau} d\tau\nn
&=& \frac{1}{2\pi} \int \gamma_{\alpha\beta}(\Omega) \left[\pi \delta(\omega-\Omega)+\frac{\ii}{\omega-\Omega}\right] d\Omega\nn
&=& \frac{1}{2}\gamma_{\alpha\beta}(\omega) + \frac{\ii}{2\pi} {\cal P} \int \frac{\gamma_{\alpha\beta}(\Omega)}{\omega-\Omega} d\Omega\,,
\eea
where we have inserted the the Fourier transform of the Heaviside-$\Theta$ function.
These principal value integrals can in principle be evaluated numerically, but for a Lorentzian tunneling rate we may also obtain an analytic solution in terms of 
Polygamma functions (not shown for brevity).
For flat tunneling rates, this Lamb shift contribution diverges logarithmically.
For example, for a Lorentzian tunneling rate $\Gamma_\alpha^{(1)}(\omega)$ we would get for large bandwidth $\delta_\alpha$ (compare e.g. App. C of Ref.~\cite{schaller2009b})
\bea\label{EQ:half_sidedexplicit}
\Gamma_{12}(\omega) &\approx& \frac{1}{2} \Gamma_{\rm L} [1-f_{\rm L}(+\omega)]\\
&&+ \ii \frac{\Gamma_{\rm L}}{2\pi} \left[\ln \frac{2\pi}{\beta_{\rm L} \delta_{\rm L}} + \Re \Psi\left(\frac{1}{2} + \ii \frac{\beta_{\rm L}(\omega-\mu_{\rm L})}{2\pi}\right)\right]\,,\nn
\Gamma_{21}(\omega) &\approx& \frac{1}{2} \Gamma_{\rm L} f_{\rm L}(-\omega)\nn
&&+ \ii \frac{\Gamma_{\rm L}}{2\pi} \left[\ln \frac{2\pi}{\beta_{\rm L} \delta_{\rm L}} + \Re \Psi\left(\frac{1}{2} + \ii \frac{\beta_{\rm L}(\omega+\mu_{\rm L})}{2\pi}\right)\right]\,,\nonumber
\eea
where $\Psi(x)$ denotes the Digamma function, and similar for the right-associated rates.
However, since in our approach the transformed TQD has flat tunneling rates $\Gamma_\alpha^{(1)}(\omega) = 2 \delta_\alpha$, we have to consider the 
limit $\delta_\alpha \to \delta_{\rm cutoff} \to \infty$ and $\Gamma_\alpha \to 2 \delta_\alpha$ in the above equations.
Numerically, we see that the currents and steady states do not depend on the cutoff width $\delta_{\rm cutoff}$.

%%%%%%%%%%%%%%%%%%%%%%%%%%%%%%%%%%%%%%%%%%%%%%%%%%%%%%%%%%%%%%%%%%%%%%%%%%%%%%%%%%%%%%%%%%%%%%%%%%%
%%%%%%%%%%%%%%%%%%%%%%%%%%%%%%%%%%%%%%%%%%%%%%%%%%%%%%%%%%%%%%%%%%%%%%%%%%%%%%%%%%%%%%%%%%%%%%%%%%%
\section{SQD efficiency}\label{SEC:sqd_efficiency}

To evaluate the efficiency of information conversion of an SQD treatment, we can use earlier results~\cite{esposito2012a}.
The entropy flow term for the rate Eq.~(\ref{EQ:ratematrix_feedback}) becomes 
\bea
\dot{S}_{\rm e} &=& \sum_\alpha \sum_{mm'} W_{mm'}^{(\alpha)} P_{m'} \ln \frac{W_{m'm}^{(\alpha)}}{W_{mm'}^{(\alpha)}}\nn
&=& \sum_\alpha \beta_\alpha \left(I_{\rm E}^{(\alpha)} - \mu_\alpha I_{\rm M}^{(\alpha)}\right) + \sum_\alpha I_{\rm M}^{(\alpha)} \ln \frac{\Gamma_\alpha^{\rm F}}{\Gamma_\alpha^{\rm E}}\nn
&=& \sum_\alpha \beta_\alpha \dot{Q}^{(\alpha)} + I_{\rm M}^{({\rm L})} \ln\frac{\Gamma_{\rm L}^{\rm F} \Gamma_{\rm R}^{\rm E}}{\Gamma_{\rm L}^{\rm E} \Gamma_{\rm R}^{\rm F}}\,.
\eea
At steady state, this becomes the negative entropy production rate, and the second law reads
\bea
-\sum_\alpha \beta_\alpha \dot{Q}^{(\alpha)} - I_{\rm M}^{({\rm L})} \ln\frac{\Gamma_{\rm L}^{\rm F} \Gamma_{\rm R}^{\rm E}}{\Gamma_{\rm L}^{\rm E} \Gamma_{\rm R}^{\rm F}} \ge 0\,,
\eea
such that an efficiency of information conversion is -- in the region of $P_{\rm el}\ge 0$ --  
given by (we assume equal temperatures and in the SQD treatment both energy and matter currents are conserved)
\bea
\eta = \frac{\beta P_{\rm el}}{I_{\rm M}^{({\rm L})} \ln \frac{\Gamma_{\rm L}^{\rm E} \Gamma_{\rm R}^{\rm F}}{\Gamma_{\rm L}^{\rm F} \Gamma_{\rm R}^{\rm E}}} = \frac{-\beta(\mu_{\rm L}-\mu_{\rm R})}{\ln \frac{\Gamma_{\rm L}^{\rm E} \Gamma_{\rm R}^{\rm F}}{\Gamma_{\rm L}^{\rm F} \Gamma_{\rm R}^{\rm E}}}\,.
\eea
Using the numerical values in the figures in the main article, this is way larger than the efficiency in Fig.~\ref{FIG:power_vs_coupling} in the weak-coupling limit.

%%%%%%%%%%%%%%%%%%%%%%%%%%%%%%%%%%%%%%%%%%%%%%%%%%%%%%%%%%%%%%%%%%%%%%%%%%%%%%%%%%%%%%%%%%%%%%%%%%%
%%%%%%%%%%%%%%%%%%%%%%%%%%%%%%%%%%%%%%%%%%%%%%%%%%%%%%%%%%%%%%%%%%%%%%%%%%%%%%%%%%%%%%%%%%%%%%%%%%%
\section{Second law with included detector}\label{APP:repeated_interactions}

One can also derive the second law by considering the implementation of the measurement apparatus in more detail.
Then, it is not necessary to perform an average over different trajectories.
We note that the treatment here fits in the framework of repeated interactions~\cite{strasberg2017a}, where the 
QPC measurements and external feedback operations assume the role of the units, which however are subject to a nonequilibrium environment.
It is sufficient to remain at the level of the average density matrix $\rho_{\rm SD}$ involving TQD system and detector and the corresponding
reduced density matrices of system $\rho_{\rm S} = \ptrace{D}{\rho_{\rm SD}}$ and detector $\rho_{\rm D} = \ptrace{S}{\rho_{\rm SD}}$, respectively.
We will consider a finite measurement and control times $\Delta t$, $\Delta \tau$, dropping however for brevity their dependence 
in the stroboscopic stationary state.

Right before the measurement, the joint density matrix is given by
\bea
\rho_{\rm SD}^{(0)} = \bar\rho \otimes \ket{0}\bra{0}\,,
\eea
and the mutual information $I \equiv S_{\rm vN}(\rho_{\rm S}) + S_{\rm vN}(\rho_{\rm D}) - S_{\rm vN}(\rho_{\rm SD})\ge 0$ of this
state actually vanishes.

As the first part of the measurement, we let TQD system and QPC interact during the time-interval $\Delta \tau$, leading to the joint density matrix
\bea
\rho_{\rm SD}^{(1)} = \sum_n \left({\cal M}_n \bar\rho\right) \otimes \ket{n}\bra{n}\,,
\eea
where the measurement superoperators are defined in Eq.~(\ref{EQ:povm_measurement}).
We note that the joint entropy of this state is exactly given by the sum of the Shannon-entropy of the detector and the averaged entropy of the system~\cite{nielsen2000}
$S_{\rm vN}(\rho_{\rm SD}^{(1)}) = -\sum_n P_n \ln P_n + \sum_n P_n S_{\rm vN}(\frac{{\cal M}_n \bar\rho}{P_n})$, where as before $P_n = \trace{{\cal M}_n \bar\rho}$.
Also, we can calculate the reduced density matrices $\rho_{\rm S}^{(1)} = \sum_n {\cal M}_n \bar\rho = e^{{\cal L}_{\rm dt}(0)\Delta\tau}\bar\rho$ and
$\rho_{\rm D}^{(1)} = \sum_n P_n \ket{n}\bra{n}$, and the mutual information between system and detector becomes
\bea
I^{(1)} = S_{\rm vN}\left(e^{{\cal L}_{\rm dt}(0) \Delta\tau}\bar\rho\right) - \sum_n P_n S_{\rm vN}\left(\frac{{\cal M}_n \bar\rho}{P_n}\right)\,.
\eea
We can confirm its non-negativity by inequality~(\ref{EQ:ent_inequality}).

During control, we apply the conditional evolution, leading to
\bea
\rho_{\rm SD}^{(2)} = \sum_n \left( e^{{\cal L}_n (\Delta t - \Delta\tau)} {\cal M}_n \bar\rho\right) \otimes \ket{n}\bra{n}\,.
\eea
The entropy of this state can also be additively decomposed into the Shannon entropy of the detector and the averaged system entropy
$S_{\rm vN}(\rho_{\rm SD}^{(2)}) = -\sum_n P_n \ln P_n + \sum_n P_n S_{\rm vN}(\frac{e^{{\cal L}_n (\Delta t - \Delta\tau)} {\cal M}_n \bar\rho}{P_n})$.
By construction, the reduced density matrices become $\rho_{\rm S}^{(2)} = \bar\rho$ and $\rho_{\rm D}^{(2)} = \sum_n P_n \ket{n}\bra{n}$, such that 
their mutual information is now
\bea
I^{(2)} = S_{\rm vN}(\bar\rho) - \sum_n P_n S_{\rm vN}\left(\frac{e^{{\cal L}_n (\Delta t - \Delta\tau)} {\cal M}_n \bar\rho}{P_n}\right)\,.
\eea
for which we can also confirm the non-negativity by inequality~(\ref{EQ:ent_inequality}).
Thus, not all of the mutual information is used to perform the feedback operation.

Finally, we reset the detector to its initial value by $\rho_{\rm SD}^{(3)} = \sum_m \ket{0}\bra{m} \rho_{\rm SD}^{(2)} \ket{m}\bra{0}$, closing the loop.
In terms of units, we would replace the old unit by a new one.
When the detector is reset, we discard the mutual information $I^{(2)}$, which explains the detrimental effect on performance. 
Explicitly, this resetting yields 
\bea
\rho_{\rm SD}^{(3)} = \left(\sum_n e^{{\cal L}_n (\Delta t - \Delta\tau)} {\cal M}_n \bar\rho\right) \otimes \ket{0}\bra{0} = \rho_{\rm SD}^{(0)}\,,
\eea
where we have used that we operate at (stroboscopic) steady state.
Accordingly, also the entropies must be the same after one feedback cycle.

During control, the detector does not change, and as we have a conventional evolution, we have for the change of total entropies~\cite{esposito2010b}
\bea
\Delta S_{\rm ct} &\equiv& S_{\rm vN}(\rho_{\rm SD}^{(2)})-S_{\rm vN}(\rho_{\rm SD}^{(1)})\nn
&=& \Delta_\ii S_{\rm ct} + \sum_{\alpha\in\{\rm L,R\}} \beta_\alpha \Delta Q^{(\alpha)}\,,
\eea
with irreversible entropy production $\Delta_\ii S_{\rm ct} \ge 0$ and heat transfers from the reservoirs $\Delta Q^{(\alpha)}$.
The measurement-associated contributions can be separated into the buildup of system-detector correlations and their removal when resetting the detector.
From the explicit expressions for the entropies we also have
\bea
-\Delta S_{\rm ct} &=& \sum_n P_n \Big[S_{\rm vN} \left(\frac{{\cal M}_n \bar\rho}{P_n}\right)\nn
&&- S_{\rm vN}\left(e^{{\cal L}_n(\Delta t-\Delta \tau)} \frac{{\cal M}_n \bar\rho}{P_n}\right)\Big]\nn
&=& \sum_n P_n \left[S_{\rm vN} \left(\frac{{\cal M}_n \bar\rho}{P_n}\right) - S_{\rm vN}(\bar\rho)\right]\nn
&&+ S_{\rm vN}(\bar\rho)\nn
&&- \sum_n P_n S_{\rm vN}\left(e^{{\cal L}_n(\Delta t-\Delta \tau)} \frac{{\cal M}_n \bar\rho}{P_n}\right)\nn
&=& \Delta S_{\rm ms} + S_{\rm vN}(\bar\rho)\nn
&&- \sum_n P_n S_{\rm vN}\left(e^{{\cal L}_n(\Delta t-\Delta \tau)} \frac{{\cal M}_n \bar\rho}{P_n}\right)\,,
\eea
where we have used Eq.~(\ref{EQ:average_measurement_entropy}).
Comparing the two expressions above and re-arranging eventually yields
\bea
- \beta_{\rm L} \Delta Q^{({\rm L})} - \beta_{\rm R} \Delta Q^{({\rm R})} - \Delta S_{\rm ms} = \Delta_\ii S_{\rm ct} + S_{\rm vN}(\bar\rho)\nn
- \sum_n P_n S_{\rm vN}\left(e^{{\cal L}_n(\Delta t-\Delta \tau)} \frac{{\cal M}_n \bar\rho}{P_n}\right) \ge 0\,,\qquad
\eea
which upon invoking Eq.~(\ref{EQ:ent_inequality}) yields the same second law~(\ref{EQ:second_law1}) as in the main manuscript.
We see that the last two terms on the r.h.s. correspond
to the mutual information $I^{(2)}$ that is discarded in the resetting of the detector.
Finally, we also mention that we can express the average of the measurement entropy change as
\bea
\Delta S_{\rm ms} &=& S_{\rm vN}(\rho_S^{(1)}) - S_{\rm vN}(\bar\rho) - I^{(1)}\nn
&=& S_{\rm vN}(\rho_{\rm SD}^{(1)}) - S_{\rm vN}(\rho_{\rm SD}^{(2)}) - I^{(2)}\,,
\eea
which demonstrates that it is not only the mutual information gathered during the measurement which bounds the performance, 
but also how much of it is actually used during the feedback.

\end{document}